\documentclass[journal]{vgtc}                     

\onlineid{1036}

\vgtccategory{Research}
 
\vgtcpapertype{algorithm/technique}

\title{Adaptively Placed Multi-Grid Scene Representation Networks \\ for Large-Scale Data Visualization}

\author{%
  \authororcid{Skylar W. Wurster}{0000-0001-6685-615X},
  Tianyu Xiong, Han-Wei Shen, Hanqi Guo, and Tom Peterka
}

\authorfooter{
  \item
  	Skylar W Wurster is with The Ohio State University.
  	E-mail: wurster.18@osu.edu
}

\abstract{%
  Scene representation networks (SRNs) have been recently proposed for compression and visualization of scientific data. However, state-of-the-art SRNs do not adapt the allocation of available network parameters to the complex features found in scientific data\textcolor{black}{, leading to a loss in reconstruction quality}. We address this shortcoming with an \textcolor{black}{adaptively placed multi-grid SRN (APMGSRN)} and \textcolor{black}{propose a domain decomposition training and inference technique for accelerated parallel training on multi-GPU systems}. \textcolor{black}{We also release} an open-source neural volume rendering application that allows plug-and-play rendering with any PyTorch-based SRN. Our proposed APMGSRN architecture uses \textcolor{black}{multiple} spatially adaptive feature grids \textcolor{black}{that learn where to be placed within the domain} to dynamically allocate more neural network resources where error is high in the volume, improving state-of-the-art reconstruction accuracy of SRNs for scientific data without requiring expensive octree refining, pruning, and traversal like previous adaptive models. In our domain decomposition approach for representing large-scale data, we train an set of \textcolor{black}{APMGSRNs} in parallel \textcolor{black}{on separate bricks of the volume} to reduce training time while avoiding overhead necessary for an out-of-core solution for volumes too large to fit in GPU memory. After training, the lightweight SRNs are used for realtime neural volume rendering in our open-source renderer, where arbitrary view angles and transfer functions can be explored. A copy of this paper, all code, all models used in our experiments, and all supplemental materials and videos are available at \url{https://github.com/skywolf829/APMGSRN}.
}


\graphicspath{{figs/}{figures/}{pictures/}{images/}{./}} 

\usepackage{tabu}                      
\usepackage{booktabs}                  
\usepackage{lipsum}                    
\usepackage{mwe}                       
\usepackage{amsmath}  
\usepackage{algorithm}
\usepackage{algpseudocode} 
\usepackage{multirow} 
\usepackage[normalem]{ulem} 
\useunder{\uline}{\ul}{}
\usepackage{wrapfig}

\usepackage{mathptmx}                  

\begin{document}


\firstsection{Introduction}
\maketitle

Scene representation networks (SRNs) are compact neural networks that map input coordinates to output scalar field values \cite{sitzmann19_siren, mildenhall20_nerf, liu20_nsvf, Reiser21_kilonerf, takikawa21_nglod}.
SRNs use a small footprint on disk (data reduction between $10-10000\times$) while being efficient to evaluate with random access.
With these benefits, SRNs have been used for compression \cite{Lu21_neurocomp, Weiss22_fvsrn} and volume rendering \cite{Weiss22_fvsrn, Wu22_ngpscivis} for scientific data up to sizes of 1TB.

Despite SRNs' popularity, there are \textcolor{black}{two} shortcomings of current state-of-the-art SRNs for scientific data visualization.
First, the SRN architectures used in recent approaches are not designed to adapt network resources to more important regions, and make the inherit assumption that the volume it will represent has uniform complexity, leading to inefficient use of the network parameters in homogeneous regions and limiting model performance.
Recently proposed SRN models have incorporated adaptive parameter allocation with spatially adaptive quadtrees and octrees that refine on the scene's geometry or features, but these approaches require significant extra storage and computation to store and search the tree structure during training \cite{Martel21_acorn, liu20_nsvf, takikawa21_nglod, yu21_plenoctrees}, taking hours or days to fit on images with fewer degrees of freedom than most 3D scientific data.
In addition to this, the tree structure itself is not directly learnable, so the trees are updated every set number of iterations with an ad-hoc method while training.
The tree structure also scales poorly with dimensionality, increasing storage and computation requirements exponentially as the number of dimensions or tree depth increases.
\textcolor{black}{Second, existing training routines for large-scale data are proposed for a single GPU with out-of-core sample streaming \cite{Wu22_ngpscivis}, which introduces I/O overhead and does not take advantage of the multiple GPUs often available on a single compute node from the supercomputers generating the large-scale data.}

\begin{figure*}[ht]
\centering
  \includegraphics[width=0.9\textwidth]{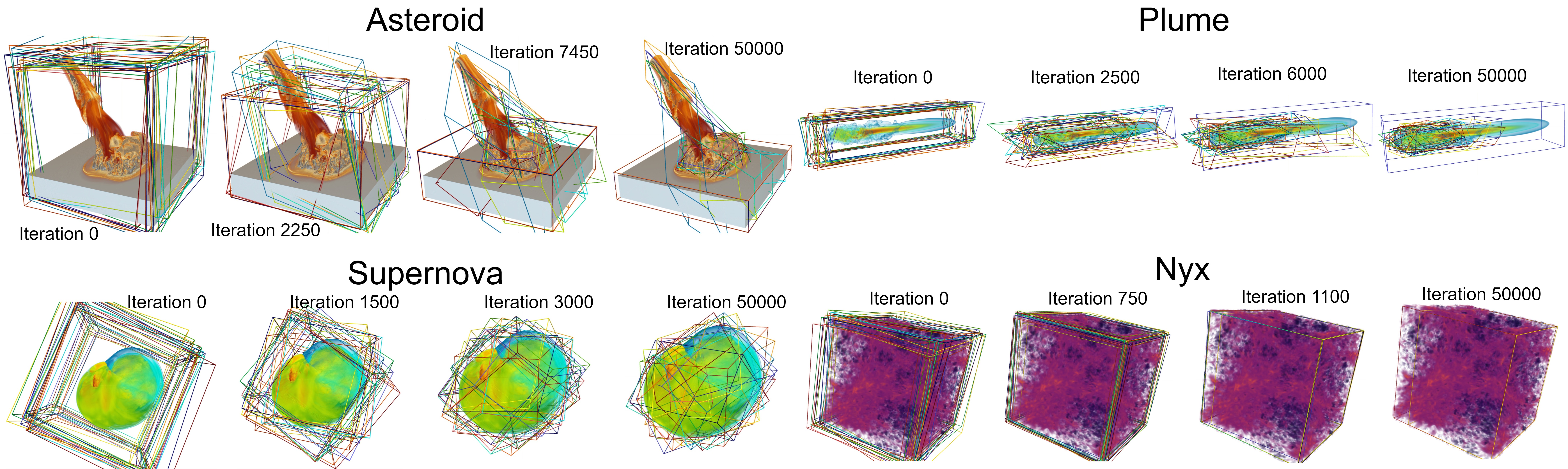}
  \caption{
  Examples of our \textcolor{black}{adaptively placed} feature grids fitting to volumes during training. The feature grids \textcolor{black}{find volume-specific regions to cover to maximize reconstruction accuracy}.
  Video of evolution of grids during training available in supplemental materials.
  }
  \label{fig:teaser}
\end{figure*}

In this work, we address the \textcolor{black}{two} limitations mentioned above with a novel SRN model and domain decomposition training routine for fitting large-scale data.
First, we address the lack of adaptivity in state-of-the-art SRNs for scientific data with a novel SRN architecture called \textcolor{black}{adaptively placed multi-grid SRN (APMGSRN)}.
Instead of using trees as \textcolor{black}{the} adaptive data model in our network, APMGSRN uses a set of spatially adaptive feature grids, shown in \autoref{fig:teaser}, whose extents are defined by learned transformation matrices which transform global space into local feature space, where the grid is defined as the unit cube $[-1,1]^3$.
To guide the grids to cover spatial regions where the model has relatively higher error, we develop a custom \textit{feature density loss}, that calculates the relative entropy from the current feature grid density to a target feature density, where the assumption is higher feature density improves reconstruction (shown by other feature grid networks \cite{Weiss22_fvsrn, Chen22_tensorf, takikawa21_nglod, yu21_plenoctrees}).
This alone is not enough to train the feature grids though, as the density of a grid is a step function that has a gradient of 0 everywhere.
Therefore, we approximate the feature density with a differentiable function that closely matches the step function called a \textit{flat-top gaussian}.
We also develop a specific training routine for the feature grid's transformation matrices with delayed start and early stopping to improve accuracy and converge quicker.
Our adaptive grid architecture does not require expensive tree searching or pruning like other adaptive models, and dynamically allocates more neural network resources to regions of higher error for any volume, improving state-of-the-art SRN performance by using network parameters efficiently.

Using our APMGSRN architecture as a building block, we \textcolor{black}{develop a \textit{domain decomposition}} training strategy that fits a large-scale volume in a model-parallel fashion.
Our domain decomposition training approach divides data in the volume on a grid of bricks, and trains one SRN per block of data.
The blocks of data fit in GPU memory, so there are no inefficiencies from an out-of-core data sampling method.
We also assume that the large-scale data being fit are \textcolor{black}{often} generated by machines equipped with multiple GPUs per node, so we use all available GPUs to train multiple networks in parallel, reducing total training time.
Inference in this \textcolor{black}{set} of models is more complicated now, since a search is necessary to find which model was trained on the spatial domain for each point being queried.
To accelerate inference \textcolor{black}{in a domain decomposition model}, we use a hash function that maps spatial coordinates to the hashtable entries for the correct model to use for inference in parallel.
Not only do domain decomposition models allow fitting \textcolor{black}{a 450 GB volume} in under 7 minutes \textcolor{black}{on 8 GPUs}, but they also increase reconstruction accuracy over a single model at the same number of total model parameters.


In summary, our contributions are \textcolor{black}{twofold}:

\begin{itemize}
    \item A novel SRN architecture called \textcolor{black}{adaptively placed multi-grid SRN (APMGSRN)} with adaptive feature grids that localize trainable network parameters on regions of the volume with high error during training
    \item A domain decomposition training \textcolor{black}{and inference} strategy that trains multiple SRNs in parallel for fitting large-scale data \textcolor{black}{quicker on capable machines}
\end{itemize}

Our APMGSRN architecture and domain decomposition modelling technique are evaluated on several scientific datasets ranging from volumes of size $128^2\times512$ (32MB to store) up to $10240\times1536\times7680$ (450GB to store).
We compare the reconstruction quality of our proposed APMGSRN architecture (single model, not \textcolor{black}{decomposed}) with other state-of-the-art models, and demonstrate that our adaptive feature grids improve reconstruction quality over state-of-the-art at similar model sizes.
We further evaluate the performance of our \textcolor{black}{domain decomposition strategy} by fitting two datasets that are 250 GB and 450 GB.
Lastly, we evaluate the \textcolor{black}{rendering} performance of our \textcolor{black}{proposed model compared with other state-of-the-art SRNs in our open-source renderer, which we release with our code to offer an easy to use, plug-and-play, neural volume renderer for INRs trained on scientific data}.
We provide all code for our model, training, and neural volume renderer on GitHub: \url{https://github.com/skywolf829/APMGSRN}, along with installation instructions and supplemental videos.

\section{Related Works}

As our method is primarily related to scene representation networks, we review related literature covering SRN architectures and training, examining applications in the computer vision and sci-vis domains.
\textcolor{black}{Our work is part of a larger domain of research called DL4SciVis, for which Wang and Han provide a comprehensive survey \cite{wang22_dl4scivis}}.

\subsection{Scene representation networks}

Scene representation networks, also called implicit neural representations (INRs) or coordinate networks, are networks that encode a given scene with the weights of the neural network, such that input coordinates are mapped to output values.
In the context of computer vision, SRNs can learn to represent images \cite{sitzmann19_siren, Martel21_acorn}, signed distance functions \cite{sitzmann19_siren, Martel21_acorn}, or radiance fields \cite{mildenhall20_nerf, Reiser21_kilonerf, yu21_plenoctrees}.
In the context of scientific data and visualization, SRNs have been used to model 3D scalar fields and time-varying scalar fields \cite{Lu21_neurocomp, Weiss22_fvsrn, Wu22_ngpscivis}.
We broadly classify the architecture of SRN models into two categories: fully connected and grid-based encoding.

\begin{figure}[ht]
\centering
  \includegraphics[width=1.0\columnwidth]{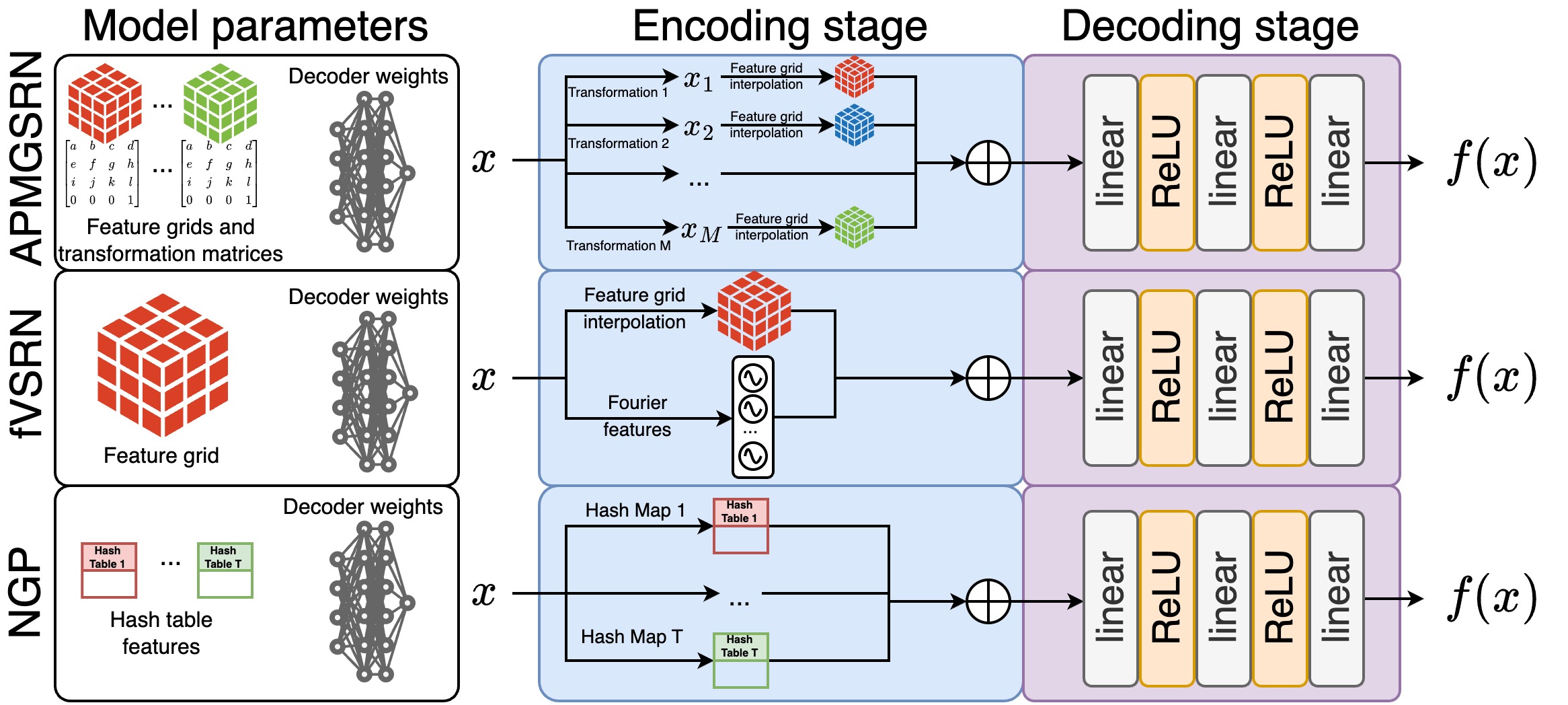}
  \caption{
    Comparison of our APMGSRN with other state-of-the-art models fVSRN \cite{Weiss22_fvsrn} and NGP \cite{Muller22_ngp}.
    The $\oplus$ operator represents concatenation.
  }
  \label{fig:modelcomparison}
\vspace{-22pt}  
\end{figure}

\begin{figure*}[ht]
\centering
  \includegraphics[width=0.8\textwidth]{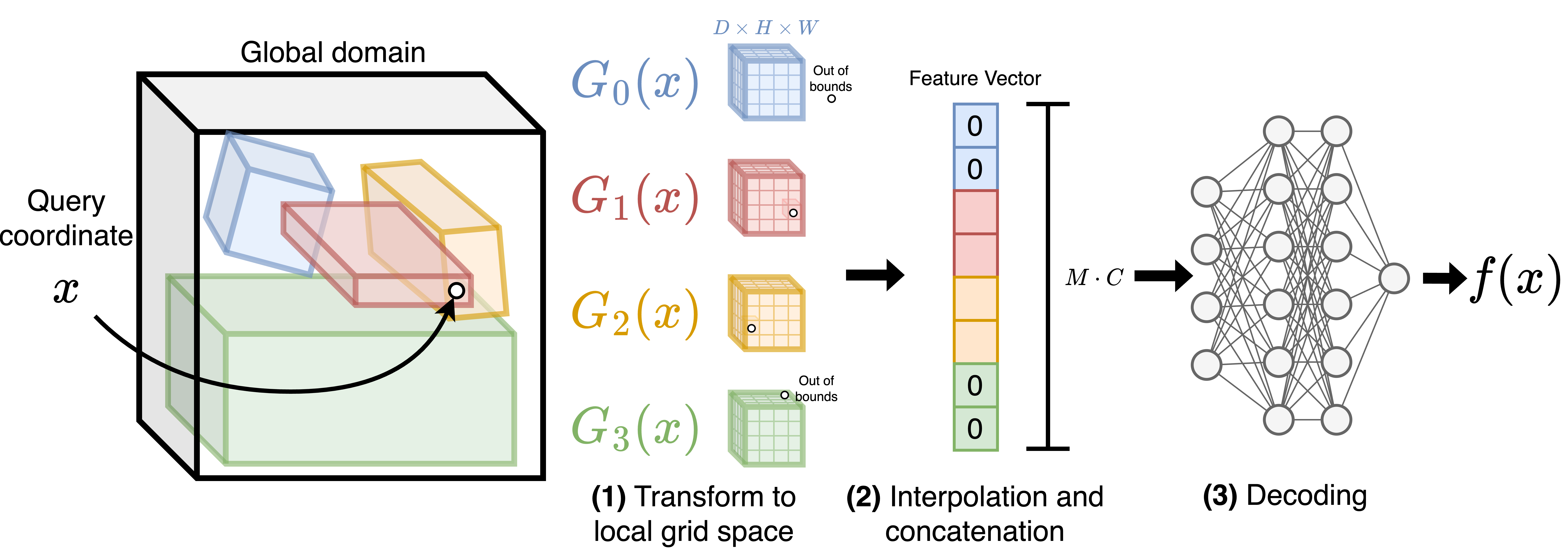}
  \caption{
  An overview of the APMGSRN architecture. 
  In \textbf{(1)}, a query coordinate is transformed into each of the $M$ grid's local coordinate systems learned by the transformation matrices, where grid local extents are assumed to be $[-1, 1]^3$.
  \textbf{(2)}, the local coordinates are used to trilinearly interpolate within each feature grid (of resolution $D \times H \times W$) to obtain the corresponding feature for each grid.
  If the coordinate is out of bounds, zeros are returned instead.
  The resulting $M \cdot C$ features are concatenated into a feature vector.
  Steps (1) and (2) are considered the encoding process.
  In step \textbf{(3)}, the feature vector is decoded in a shallow MLP for the final output value $f(x)$.
  }
  \label{fig:architecture}
\vspace{-18pt}  
\end{figure*}

\textbf{Fully connected SRNs.}
Fully connected SRNs use only linear layers with activations between them to map an input coordinate to output value, making them slow to train and perform inference on.
SIREN \cite{sitzmann19_siren} is a fully connected SRN with \textcolor{black}{sinusoidal activations to fit data accurately.
Fourier features by Tancik et al. \cite{tancik20_fourier} shows that Fourier encoding of input coordinates can help INR accuracy.}
NeRF \cite{mildenhall20_nerf} is a fully connected SRN for modelling a radiance field from multi-view images of a scene.
AutoInt \cite{lindell21_autoint} builds on NeRF by requiring far less model inferences during neural volume rendering compared to NeRF by exploiting  that the gradient of a SIREN network is another SIREN network that shares the same weights as the original network.
Then the ``gradient'' network can be trained as a NeRF, resulting in the original ``integral'' network learning the integration of colors and density through rays in space.
Lu et al. \cite{Lu21_neurocomp} use a SIREN-based architecture with residual connections to model 3D volumetric data in a compressed format.
\textcolor{black}{Höhlein et al. \cite{Höhlein22_vrncompression} study the capabilities of INRs for compressing meteorlogical ensemble data}.
Han and Wang \cite{han22_coordnet} propose CoordNet, a single implicit model based on the residual siren architecture of Lu et al. \cite{Lu21_neurocomp} that is used for spatiotemporal super resolution and novel view synthesis.
\textcolor{black}{Tancik et al. \cite{tancik2022_blocknerf} propose Block-NeRF, which uses multiple NERF models representing disjoint subsections of blocks of a region of San Francisco. 
This is similar to our approach, but models represent geometry (learned from images) within a specified radius from the model's center with large overlap with other models, while our approach divides the domain with rectangular bricks as it fits the voxel grid representation better.}

\textbf{Grid-based encoding SRNs.}
In contrast to fully connected SRNs, grid-based encoding SRNs move a large majority of the network parameters to an efficient encoding scheme that transforms the input coordinates to a high-dimensional feature space.
SRNs with grid-based encoding schemes train and infer faster than their fully connected counterparts due to the efficient encoding and limited fully connected operations.
Weiss et al. \cite{Weiss22_fvsrn} create fVSRN, an SRN that models 3D scientific data using a feature-grid encoder which places a low-resolution feature grid over the data domain and interpolates within the feature grid to obtain a feature vector for decoding.
Yu et al. \cite{yu21_plenoctrees} accelerate a pre-trained NeRF model with spherical harmonics and an octree data structure for empty-space skipping, enabling realtime rendering of NeRF models.
Liu et al. \cite{liu20_nsvf} propose neural sparse voxel fields, which uses a coarse 3D grid of voxels that are pruned and split at checkpoints during training to reduce the training time of NeRF with improved accuracy.
\textcolor{black}{Genova et al. \cite{genova2020_local} learn an implicit representation of a mesh from depth images using a dense grid and many local deep implicit functions}.
Takikawa et al. \cite{takikawa21_nglod} propose NGLOD, which learns an octree data structure for realtime rendering of an SRN fitting a signed-distance function.
Martel et al. \cite{Martel21_acorn} create an adaptive coordinate network that uses quadtrees/octrees during training, updated by solving an integer linear programming optimization problem every set number of iterations.
Chen et al. \cite{Chen22_tensorf} use tensor (de)composition to reduce the space complexity of 3D feature grids with their VM decomposition of a tensor.
\textcolor{black}{M\"{u}ller} et al. \cite{Muller22_ngp} introduce hash grid encoding for neural graphics primitives (NGP), which uses random hashing to map input coordinates to features in a hash table at multiple levels of detail.
Wu et al. \cite{Wu22_ngpscivis} use the hash grid architecture\cite{Muller22_ngp} to model scientific data up to 1TB in size.
\textcolor{black}{Concurrent work by Wu et al. \cite{wu2023_hyperinr} uses a meta-learning approach that predicts the network weights of a set of models of the hash grid architecture \cite{Muller22_ngp} for novel view synthesis and neural volume rendering of temporally interpolated volumes and dynamic global shadows.}

Existing adaptive models such as ACORN \cite{Martel21_acorn}, NSVF\cite{liu20_nsvf}, and NGLOD \cite{takikawa21_nglod} are expensive to train, store, and require ad-hoc steps for pruning/refining the tree data structure. 
Additionally, the tree-based adaptivity only helps for scenes with static surfaces. 
Volumetric scientific data do not generally have static surfaces, since different surfaces arise from visualizing different isovalues.
Thus, the space-skipping and tree structures used by other methods ignore relevant regions of the scientific data.
Our approach's adaptive feature grids do not require surfaces to refine to, and is driven only by reconstruction error.
APMGSRN is conceptually an adaptive version of fVSRN that splits the single highly-parameterized feature grid into many less-parameterized feature grids, where each feature grid now has the ability to adjust its transformation within the domain to focus on high-error regions.
A visual comparison of our model and two state-of-the-art models compared with in this paper, fVSRN and NGP, is provided in \autoref{fig:modelcomparison}.

\section{Overview}

Our approach is composed of \textcolor{black}{two} components: (1) a novel SRN architecture called APMGSRN, and (2) a domain decomposition training and inference method to train on large-scale data. 
In \autoref{APMGSRN}, we detail the APMGSRN architecture, a SRN model that adaptively learns where multiple feature grids should be spatially located while training.
In \autoref{domaindecompositionSRN}, we explain the \textcolor{black}{domain decomposition approach}, which dissects a volume into a 3D grid of networks that are each trained on their own local region.

\section{\textcolor{black}{Adaptively placed multi-grid scene representation network}}
\label{APMGSRN}

Recent state-of-the-art SRNs have found that using explicitly defined feature grids within the model reduces training and inference times \cite{Weiss22_fvsrn, Chen22_tensorf, Muller22_ngp}.
In these feature grid-based models, a feature grid is interpolated at an input spatial coordinate to retrieve a feature vector, which is fed through a shallow multi-layer perceptron (MLP) to obtain the final output value.
Tree structures have been added to SRN models to support adaptive allocation of network resources for specific regions \cite{liu20_nsvf, Martel21_acorn, yu21_plenoctrees}, but the approaches are ad-hoc, require hours to days to train, and have significant training/storage overhead to manage the tree structure.
Additionally, the cost of the tree structures scale exponentially as the number of dimensions or tree depth increases.

Our method improves adaptivity within SRNs with a novel \textcolor{black}{adaptively placed multi-grid scene representation network (APMGSRN)}, depicted in \autoref{fig:architecture}.
Instead of a tree structure, APMGSRN uses a set of multiple spatially adaptive feature grids (shown in \autoref{fig:teaser}), each described by a learnable transformation matrix, for coordinate encoding before being decoded by a shallow MLP, as described in \autoref{adaptivefeaturegrids}.
Since a basic reconstruction loss is not enough to learn the transformation matrices properly, \autoref{learningmatrices} describes our method to learn grid positions with a feature density-based loss function.
Finally, we discuss the network training in \autoref{APMGSRNtraining}.
Our model gives strong reconstruction quality even with few model parameters thanks to the ability to transform the feature grids to fit the complexities in the data.
Since our model does not rely on a tree structure, inference remains quick and storage costs remain low.

\subsection{APMGSRN architecture}
\label{adaptivefeaturegrids}

An APMGSRN is a function $f(x)$ that maps normalized spatial coordinates $x \in [-1,1]^3$ to the scalar value at that location in space.
The model architecture is composed of an encoder $e$ and decoder $d$ such that $f(x) = d(e(x))$, depicted in \autoref{fig:modelcomparison} and \autoref{fig:architecture}\textcolor{black}{.}
Our encoder, described in \autoref{encoder}, is our \textcolor{black}{adaptively placed multi-grid encoder}, and our decoder, described in \autoref{decoder}, is a small fully-connected MLP that decodes the encoded feature vector to an output value.

\subsubsection{\textcolor{black}{Adaptively placed multi-grid encoder}}
\label{encoder}
The encoder in APMGSRN contains $M$ learnable feature grids of resolution $D\times H \times W$ with $C$ channels, represented as a tensor $F$ with shape $[M, C, D, H, W]$.

\textbf{Transformation to local space.}
In our model, the spatial extents of the $i$-th feature grid is defined by a $4\times4$ transformation matrix $G_i$, which transforms global-space coordinates to local-space coordinates according to the following:
\begin{equation}
    \label{transformationMultiplication}
    \begin{bmatrix} p_x^l \\ p_y^l \\ p_z^l \\ 1 \end{bmatrix} = \begin{bmatrix}
    G_{i,0,0} & G_{i,0,1} & G_{i,0,2} & G_{i,0,3} \\ 
    G_{i,1,0} & G_{i,1,1} & G_{i,1,2} & G_{i,1,3} \\ 
    G_{i,2,0} & G_{i,2,1} & G_{i,2,2} & G_{i,2,3} \\ 
    0 & 0 & 0 & 1
    \end{bmatrix}
    \begin{bmatrix} p_x^g \\ p_y^g \\ p_z^g \\ 1 \end{bmatrix},
\end{equation}
where $G_{i,r,c}$ is the $r$-th row and $c$-th column of the transformation matrix, $p_x^g,p_y^g,p_z^g$ is the global coordinate, and $p_x^l,p_y^l,p_z^l$ is the resulting local coordinate.
Conceptually, the first 3 rows and columns of $G_i$ are responsible for scale, rotation, and shearing of the grid, while the 4th column is responsible for translation.
The grid's local extents are assumed to be $[-1, 1]^3$, and global coordinates for the 8 corners of a grid can be determined by calculating the inverse of $G_i$ and transforming the coordinates of the local extents to global space.
For shorthand, we consider $G_i(x_g)$ to be a function mapping a 3D global coordinate $x_g$ to the coordinate in grid $i$'s local space according to \autoref{transformationMultiplication}.

\textbf{Encoding.}
To encode an input global coordinate, the coordinate is transformed into each grid's local coordinate system via the defining transformation matrices.
Then, each 3D local coordinate $x_l$ is used to perform trilinear interpolation within each feature grid:
\begin{equation}
    \label{singleGridEncoding}
    e_i(x_l)= 
    \begin{cases}
        \texttt{interp}(F_i,x_l)& \text{if } -1 \leq x_l \leq 1\\  
        0,              & \text{otherwise}
    \end{cases}
\end{equation}
where $F_i$ is the feature grid for the $i$-th grid and $\texttt{interp}(F_i,x_l)$ performs trilinear interpolation within $F_i$ at point $x_l$ assuming the grid has vertices corner-aligned at the extents of $[-1, 1]^3$.
When the input coordinate is not within the feature grid, the feature returned is 0 for each of the expected $C$ channels.
The encoding step for a single grid as described in \autoref{singleGridEncoding} is performed for each grid, with results concatenated together.  
From global space, the encoding is calculated as:
\begin{equation}
    \label{fullencoding}
    e(x_g) = e_0(G_0(x_g)) \oplus e_1(G_1(x_g)) \oplus... \oplus e_{M-1}(G_{M-1}(x_g))
\end{equation}
where $\oplus$ is the concatenation operator.
The result is a feature vector $y = e(x_g)$ with $C\cdot M$ features, where $C$ is the number of channels and $M$ is the number of grids.

\subsubsection{Decoding}
\label{decoder}

Our decoder $d$ is a shallow 2-layer MLP $m$ with no bias terms used and 64 neurons per layer, which is lightweight and efficient.
After each layer besides the last layer, ReLU activation is used as a nonlinearity.
We use the Tiny-CUDA-NN package \cite{muller21_tcnn} for tensor-core accelerated decoding, which reduces training time by about 10\% compared to a pure PyTorch decoder.

\textbf{Scaling}.
Unlike other data formats that SRNs represent such as images,  radiance fields, and distance fields, scientific data values are generally not bounded, posing a challenge for the decoder to accurately represent the unknown data range.
We tackle this with a preprocessing step before training that identifies the minimum and maximum values for the volume the network will represent and saving those along with the network.
Then, the decoding is calculated as $d(y) = m(y) \cdot (\texttt{max}-\texttt{min}) + \texttt{min}$ where $m$ is the MLP in the decoder, $\texttt{min}$ is the saved minimum value, and $\texttt{max}$ is the saved maximum value.
This formulation allows us to avoid numerical issues by scaling the network output, which results in the MLP learning to output values between 0 and 1.
More specific scaling for specific data types, such as z-score-, log-, and exponential-scaling, is not tested in this paper and is left to future work.

\subsection{Learning feature grid transformation matrices}
\label{learningmatrices}

The adaptivity of our model stems from the learnable transformations for each feature grid in the encoder.
The goal during training is to localize feature grids on/around regions that have high error.
We focus more network parameters to those spatial regions with high error in order to improve reconstruction quality in that region.

While the idea is intuitive, naively attempting to learn the transformation matrices defining each grid is not possible using only a reconstruction loss (such as L1 or MSE). 
This is a direct consequence of our encoding (\autoref{singleGridEncoding}), which returns a 0 feature if a query point is outside of a grid.
This means the only gradients that will update a transformation matrix are from points that reside within that grid, so high error regions outside of a feature grid will not ``pull'' the feature grid toward it as desired.

To properly train the feature grid's transformation matrices, we introduce a feature density loss, which is a quantifiable and differentiable metric for the difference between the current feature density $\rho$, and a derived \textit{target} feature density $\rho^*$.
We use the term \textit{feature density} to describe the average number of features (from the feature grids) that exist per unit volume at some point in space.
In this section, we will discuss how we calculate the current feature density in a differentiable manner, how we derive a \textit{target feature density} that warps the current feature density to increase where the error is relatively higher and decrease where error is relatively lower, and a loss function to quantify the error between the current and target feature density which is ultimately used to update the transformation matrices during training.

\textbf{Feature density}.
The simplest way to represent feature density is by the number of grids that overlap a spatial position $x$.
\textcolor{black}{The feature density $\rho$ at a global location $x$ using this formulation} can be described by the following equation:
\begin{equation}
    \rho(x) = \sum_{i=0}^{M-1} \begin{cases}
        1& \text{if } -1 \leq G_i(x) \leq 1\\  
        0,              & \text{otherwise}
    \end{cases},
\end{equation}
where $G_i(x)$ is the function that transforms global coordinate $x$ into feature grid $i$'s local coordinate frame.
The if statement is true when it holds for each of the three dimensions of $G_i(x)$.
However, this approach does not accurately represent the \textit{density} of features at that point.
Each feature grid has the same resolution, but a feature grid with small extents will have more dense features than a feature grid with near global extents.
Instead, a better feature grid formulation is:
\begin{equation}
    \rho(x) = \sum_{i=0}^{M-1} \begin{cases}
        \det(G_{i,0:3,0:3}) & \text{if } -1 \leq G_i(x) \leq 1\\  
        0,                  & \text{otherwise}
    \end{cases},
\end{equation}
where $G_{i,0:3,0:3}$ is the top-left $3 \times 3$ matrix of $G_i$ representing scale, shear, and rotation for the grid, and $\det(\cdot)$ is the determinant of a matrix.
As the grid becomes more dense (smaller in the global domain), the determinant above increases, representing the true feature density of the grid.

Though an accurate description of the feature density of the encoder, this formulation suffers from a large drawback in that there is no gradient to change the transformation matrices $G$ when a point is outside of a grid, as the gradient of $\rho$ will be 0 everywhere.
This is a critical component for learning the transformation matrices in our method, and without it, the grid's scales, translations, etc., cannot update based on the error of points outside of their local domain.

Therefore, we use a gaussian approximation of the feature density, which is differentiable everywhere.
Specifically, we use a class of gaussian functions called \textit{flat-top} gaussians or \textit{super-gaussians}. 
A flat-top gaussian is the same as the normal gaussian equation with a $p$ term in the exponent:
\begin{equation}
    \label{flattopequation}
    g(x,p) = A \exp \left( -\left( \frac{(x-\mu)^{2p}}{2\sigma^2} \right) \right),
\end{equation}
where $A$ is a normalizing coefficient, $x$ is the location, $\mu$ is the center of the distribution, $\sigma$ is the standard deviation of the distribution, and $p$ is the strength of the flat-top.
As $p$ increases, the function's shape become more box-like. See our supplemental material at the end of this document for an example.

Using the flat-top approximation for feature density box function, we give the final form of $\rho(x)$ as:
\begin{equation}
    \rho(x) = \sum_{i=0}^{M-1} \left(\det \left(G_{i,0:3,0:3}\right) \cdot \exp \left( - \left( \sum_{d=0}^2 \left(G_i(x)_d\right)^{2p} \right)\right)\right),
\end{equation}
where $G_i(x)_d$ is the transformed coordinate in the x-, y-, and z-axis for $d=0,1,2$. 
Since the center of the local coordinate space is 0 and the standard deviation is 1, the gaussian equation does not require the $\mu$ or $\sigma$ terms in the transformed space.
This formulation is differentiable everywhere, facilitating proper training.
Experimentally, we find $p=10$ to be strong enough to match the box shape of the feature grid without being too strong so as to have exploding gradients on the ramps to and from the flat-top.

\textbf{Target feature density.}
With a way to calculate our current feature density at an arbitrary point, we now require a target feature density to steer toward during training.
Our goal is to increase $\rho(x)$ where the error is relatively high and decrease $\rho(x)$ where error is relatively low.
We formulate a target feature density for a coordinate $\rho^*(x)$ according to the following:
\begin{equation}
    \label{targetdensityequation}
    \rho^*(x) = \exp \left( \frac{\overline{h}}{h(x)+\epsilon} \log \left( \rho_{\text{scaled}}(x)+\epsilon \right) \right)
\end{equation}
where $h(x)$ is the model's error at location $x$, $\overline{h}$ is the average model error (in practice, the average error over a batch), $\rho_{\text{scaled}}(x)$ is $\rho(x)$ divided by its sum over the set (to rescale the density between 0 and 1), and $\epsilon$ is a small number to avoid dividing by zero and other numerical issues.
In language, \autoref{targetdensityequation} will logarithmically scale the current (scaled) feature density according to the \textit{relative error} at each position.
If the error is exactly the average, then the target feature density is unchanged with respect to $\rho_{\text{scaled}}$.
This logarithmic scaling assures that more importance is placed \textcolor{black}{on} increasing feature density in regions with very large relative error, and decreasing feature density in regions with very small relative error.

\textbf{Feature density loss function.}
To quantify the difference between the current feature density and the target feature density, we view this as a distribution-matching  problem.
We calculate our feature density loss over a set of coordinates $X$ as
\begin{equation}
    \label{featuredensityloss}
    \mathcal{L}_{\text{density}} = \frac{1}{|X|} \sum_{x \in X} \rho^*(x) \log \left( \frac{\rho^*(x)}{\rho_{\text{scaled}}(x)} \right),
\end{equation}
where $|X|$ is the number of coordinates in $X$.
The loss function measures the relative entropy from the current feature density to the target feature density, and is also known as the Kullback–Leibler divergence (KL divergence).
This loss function provides gradients that can effectively update the parameters of each feature grid to match the calculated target feature density.

\subsection{APMGSRN training}
\label{APMGSRNtraining}
In this section, we cover the losses, training routine, and initialization we use for APMGSRN.

\subsubsection{Loss functions.}
We use two loss functions while training - a reconstruction loss $\mathcal{L}_{\text{rec}}$ and the density loss  $\mathcal{L}_{\text{density}}$ described in \autoref{featuredensityloss}.
The reconstruction loss is the mean-squared error (MSE) between the model output and the ground truth data $\mathcal{L}_{\text{rec}} = \frac{1}{|X|} \sum_{x \in X} \left( f(x) - v(x)\right)^2$, where $X$ is a batch of coordinates, $|X|$ is the number of coordinates, $f$ is the SRN, and $v(x)$ returns the ground truth data value at spatial position $x$ via trilinear interpolation.

\subsubsection{Training routine.}
\label{trainingroutine}
Each training step has two parts: (1) update the feature grid and decoder parameters using $\mathcal{L}_{\text{rec}}$, (2) update the transformation matrices using $\mathcal{L}_{\text{density}}$.
\textcolor{black}{Specifically, $\mathcal{L}_{density}$ is the only loss providing gradients to update the transformation matrices $G$, and $\mathcal{L}_{\text{rec}}$ is the only loss providing gradients to update the feature grid values $F$ and decoder weights.}
As long as the model is training, step (1) will happen every iteration. 
However, performing step (2) during each iteration would slow down training and potentially negatively impact reconstruction quality for reasons discussed in the following paragraphs.
Therefore, we use a few techniques to minimize the number of step (2) updates.

\textbf{Delayed start.}
Since $\mathcal{L}_{\text{density}}$ is dependent on the error for the current training iteration $\mathcal{L}_{\text{rec}}$, the update to the transformation matrices will be more useful when the model has already seen enough to know where the low- and high-error regions will be.
When the model begins training, the error distribution may be random based on the network initialization, grid initialization, and data.
Therefore, we delay the training for the transformation matrices until the rest of the network has been briefly trained on the data so as to remove most of the noise from initialization.
We find that 500 iterations (with our learning rate and network sizes) is enough to provide clear details about what regions of the volume will be hard to learn, so the transformation matrices are frozen for the first 500 iterations, and are updated according to $\mathcal{L}_{\text{density}}$ after that.

\textbf{Early stopping.}
Our formulation presents a challenge for the neural network parameters in our APMGSRN in that if both training step (1) and (2) occur simultaneously, the model is training to chase a moving target.
The parameters are updated from $\mathcal{L}_{\text{rec}}$ with the assumption that the feature grids are static, but the grids are moving with each update step (2).
Fixing the feature grids as early as possible is essential for the network to fine-tune the feature grid parameters to their location in space, as well as to reduce training time spent updating the transformation matrices when they may have already converged.

While training the transformation matrices, we keep a running track of $\mathcal{L}_{\text{density}}$ each iteration, and set an early stopping flag when the 1000-iteration moving average of $\mathcal{L}_{\text{density}}$ has not reduced by 0.01\%.
Alternatively, if this early stopping criteria is not met after 80\% of the total training steps, we stop updating the transformation matrices in order to allow the other network parameters to learn given fixed feature grid positions.

A single APMGSRN model takes at most 4 minutes to train for 50k iterations with the hyperparameters experimented with in this paper.
Often, the grids converge quickly, reducing training time to as short as 40 seconds.

\subsubsection{Initialization}
\label{initialization}
With feature grid positions being crucial to the performance of our model, the initialization of the transformation matrices is a relevant factor in training speed and model accuracy.
We initialize our transformation matrices to cover a near global domain with small random shears, rotations, and translations.
The diagonals of the matrices (representing the scale of the grid) are sampled from $\mathcal{N}(1, 0.05)$, while the remaining entries in the first three rows of the transformation matrix are sampled from $\mathcal{N}(0, 0.05)$.
This initialization assures that each (relevant) entry in the transformation matrix is non-zero so that it may contribute to the final output, guaranteeing gradients can be used to update each entry of the matrix.
Additionally, the domains created with this initialization scheme are all nearly equal to the global extents, and so while fixed at the beginning of training (see \autoref{trainingroutine}), each grid is initially helping learn the global domain, providing a better approximation of which regions will be challenging to learn.

Besides our transformation matrices, we initialize our feature grids from $\mathcal{U}(-0.0001, 0.0001)$, encouraging near-zero initial guesses with small randomness as following M\"{u}ller et al. \cite{Muller22_ngp}.
Our decoder network weights are initialized following Glorot and Bengio \cite{glorot10_xavierinit}.

\section{Domain decomposition scene representation}
\label{domaindecompositionSRN}

It may not always be feasible to train a single model for some large-scale data for two reasons.
From one end, as training data become more complex with higher resolutions, larger neural networks are necessary to obtain adequate reconstruction accuracy, which means longer training times and a large model footprint on disk.
As model complexity increases, GPU memory may become a bottleneck during training or training time becomes unacceptably long.
From the other end, the large-scale data we wish to model with an SRN may not fit within GPU memory for efficient query in the training loop.
The main CPU memory can be used to host data and support on-demand data transfer to the GPU for training, but is costly due to the random sampling of data points during training.
If the CPU memory also cannot support hosting the data, then an out-of-core solution is the only option available.
Both options have been explored by Wu at al. \cite{Wu22_ngpscivis}.

Instead of a complicated and inefficient out-of-core sampling method, we take a model-parallel approach to modelling a large-scale volume.
We propose that instead of a single network representing the volume, we use a domain decomposition of models, where each model has learned a separate brick of the volume.
The models are arranged in a grid, and do not require communication during training or inference.
We discuss the data partitioning in \autoref{domaindecompositionpartitioning}, the training procedure in \autoref{domaindecompositiontraining}, and querying a \textcolor{black}{set of models} during inference in \autoref{domaindecompositioninference}.

\subsection{Data partitioning}
\label{domaindecompositionpartitioning}
To partition data into bricks for one network to be trained per brick, we use a grid of networks covering the domain. See our supplemental material for a figure illustrating this.
Ahead of training, grid resolution $I,J,K$ must be chosen, representing the number of networks for the width, height, and depth of the volume, respectively. 
We recommend picking an $I,J,K$ such that the resolution of the grid assigned to each network is roughly the same aspect ratio as the feature grids in the model.
For instance, if the feature grids are $32^3$, choose $I,J,K$ such that each model is assigned a roughly cube-shaped volume.

\textbf{Ghost cells.}
In order to mitigate boundary artifacts along adjacent network boundaries during visualization, we experiment with the use of a ghost layer of cells, which are cells that overlap between networks. 
During data partitioning, this means that the extents of a network will get extended by some number of ghost cells along each axis.
This solution does not guarantee that seams will be removed, but tends to reduce the effect of them, as the networks should have less of a difference along boundaries if they are learning beyond the actual extent.
\textcolor{black}{In our experiments in \autoref{domaindecompositionevaluation}, we test with between 1 and 16 ghost cells, for example.}
We discuss a more in-depth solution in our future work, but do not go beyond ghost cells in this paper.

\subsection{Domain decomposition training}
\label{domaindecompositiontraining}

To reduce the training time needed for the $I \cdot J \cdot K$ models, we dissect the domain and train the models in parallel across available GPUs. 
Since large-scale data are often generated by powerful machines, we assume a user is likely training on \textcolor{black}{a} compute node with multiple GPUs.
Our domain decomposition training routine pre-calculates the extents of each network within the \textcolor{black}{volume's extents} (including ghost cells) and generates a list of jobs that are assigned to the available GPUs.
When a GPU finishes training, the GPU returns to the available GPUs list, where it waits to be issued the next model to train.
When a GPU begins training, an APMGSRN model is initialized, and only the data within its assigned extents are loaded to the GPU memory.
This reduces I/O overhead and is  efficient with data formats such as NetCDF supporting parallel file access and arbitrary cropping from disk.

\textbf{Early stopping.}
Just as we employ early stopping to stop learning the transformation matrices earlier during training to converge faster, we also use early stopping on models within \textcolor{black}{the domain decomposition} while training so the GPU is freed for the next model to train quicker.
We use a plateau learning rate scheduler that detects when $\mathcal{L}_{\text{rec}}$ has not decreased by 0.01\% over 500 iterations. 
When this is triggered, the learning rate of the model's parameters is reduced by a factor of 10.
When this is triggered 3 times, we finish training the model.
This can reduce training times dramatically in regions where there may be very sparse data (down to 20 seconds per model in our experiments).

\subsection{Domain decomposition inference}
\label{domaindecompositioninference}

After training (presumably on a remote server), the neural networks can be moved to a local workstation for visualization.
Once on a local machine, all models are loaded into GPU memory.
Since each network in the domain decomposition represents a certain spatial extent, each global query location needs to be mapped to the correct model for inference, so we develop a spatial hashing function for efficient inference.

\textbf{Spatial hashing function.}
To efficiently determine what model a query location belongs to, we implement a spatial hashing function to hash input 3D global coordinates to the correct index in the array of models.
We load models into an array in x-,y-,z-dimension order.
With this order in mind, we can hash input global coordinates directly to array index.
Assuming a global spatial coordinate $p$ is in the domain $[-1,1]^3$, we calculate the grid index for $p$ with $i = \left \lfloor I\frac{p_x+1}{2} \right \rfloor, j = \left \lfloor J\frac{p_y+1}{2} \right \rfloor, k = \left \lfloor K\frac{p_z+1}{2} \right \rfloor$, where $\lfloor \cdot \rfloor$ is the integer floor operation.
If any dimension of the coordinate has exactly $1.0$ for an axis, then the result is an out of bound location, so these values need to be clamped such that $i<I, j<J, k<K$.
Since the networks are stored in a list, the 3D $i,j,k$ index for the network is flattened in C-order, giving the final hash as $i+Ij+IJk$.
With the correct network determined to perform inference with, the global coordinate can be scaled to the networks local domain for proper inference.

Our implementation loops over all networks to do evaluations for each, but a significant speedup could be achieved with further engineering effort, as shown by Reiser et al. \cite{Reiser21_kilonerf}, who develop custom CUDA code for inference within their \textcolor{black}{set} of networks for radiance fields.
We leave this to future work, but expect the inference time of large domain decomposition models to be much quicker when inference code is written in custom CUDA code that parallelizes network evaluation.

\section{Neural volume rendering}
\label{neuralvolumerendering}

\textcolor{black}{For fair comparison between the models,} we develop a Python/PyTorch\cite{NEURIPS19_pytorch}-based neural volume renderer (included with our code on GitHub) that supports our own APMGSRN as well as state-of-the-art models fVSRN \cite{Weiss22_fvsrn} and neural graphics primitives (NGP) using a hash-grid encoding \cite{Muller22_ngp}, on top of raw data volume rendering.
Any PyTorch model that can support mapping 3D coordinates to an output density can also be plugged in with minimal reconfiguration.
\textcolor{black}{We favor ease of use over raw performance with this renderer with a plug-and-play style coding requiring minimal changes to a user's PyTorch model, and hope it is useful for future research on SRNs for sci-vis.}
Our renderer supports arbitrary transfer function and view direction, and uses progressive rendering to support 60+ fps rendering for immediate feedback while panning, rotating, and zooming.
We support transfer functions exported from ParaView directly to our renderer, and interacting with the scene follows the typical 3D viewer paradigm of clicking and dragging rotating around the scene, the scroll wheel zooms, and middle-mouse clicking and dragging pans the camera.
We use 3D rendering helper functions from the nerfacc Python package \cite{Li22_nerfacc}, which provides CUDA-accelerated ray marching and compositing.

For improved interactivity, we implement a progressive rendering scheme that renders the image in a checkerboard pattern, evaluating the pixels in order of an image hierarchy.
After each progressive rendering pass, the in-progress image displayed on the screen is a bilinearly-upscaled version of the current finest-available image from the image hierarchy masked with the fully evaluated pixels.
We also offer a trade-off between interactivity and full render time with a user choice for batch size, with large batch sizes slowing down interactivity but speeding up full-frame render times, and vice versa for small batch sizes. 
Our progressive rendering keeps GPU memory minimized and quickly generates good approximations of the final image while offering interactive frame rates (30-400 fps).


\begin{table}[ht!]
\centering
\caption{
Datasets, their size on disk (in raw, single-precision floating point representation), and their descriptions, as well as quantitative results for SRN models trained to represent each dataset. 
Training type is either single (for single model) or an \textcolor{black}{decomposition} grid resolution for our domain decomposition training approach.
For domain decomposition models, average training time per model is reported since models are trained in parallel.
Best performing models in one size category are bold and underlined.
Peak signal-to-noise ratio (PSNR) is given in the data space.
}
\label{quanttable}
\resizebox{\columnwidth}{!}{%
\begin{tabular}{c|c|llll}
\multicolumn{1}{l|}{Dataset} & Model storage size & \multicolumn{1}{c}{Model} & \multicolumn{1}{c}{Training type} & \multicolumn{1}{c}{PSNR (dB) $\uparrow$} & \multicolumn{1}{c}{Training time} \\ \hline
\multirow{10}{*}{\begin{tabular}[c]{@{}c@{}}Plume\\ $512 \times 128^2$\\ 32 MB\\ Velocity magnitude of \\ particles in a solar \\ plume simulation.\\ Somewhat sparse.\end{tabular}} & \multirow{3}{*}{Small - 280 KB} & fVSRN & single & \textcolor{black}{43.924} & \textcolor{black}{58s} \\
 &  & NGP & single & 46.713 & 1m 32s \\
 &  & APMGSRN & single & {\ul \textbf{48.954}} & 52s \\ \cline{2-6} 
 & \multirow{3}{*}{Medium - 4 MB} & fVSRN & single & \textcolor{black}{49.332} & \textcolor{black}{57s} \\
 &  & NGP & single & 50.847 & 1m 7s \\
 &  & APMGSRN & single & {\ul \textbf{56.327}} & 1m 20s \\ \cline{2-6} 
 & \multirow{4}{*}{Large - 64 MB} & fVSRN & single & \textcolor{black}{52.165} & \textcolor{black}{1m 46s} \\
 &  & NGP & single & 53.282 & 2m 10s \\
 &  & APMGSRN & single & 58.134 & 2m 25s \\
 &  & APMGSRN & $4 \times 1 \times 1$ & {\ul \textbf{58.363}} & 56s \\ \hline
\multirow{10}{*}{\begin{tabular}[c]{@{}c@{}}Nyx \cite{nyx}\\ $256^3$ \\ 64 MB\\ Log density of dark matter \\ in cosmological simulation. \\ Somewhat sparse.\end{tabular}} & \multirow{3}{*}{Small - 280 KB} & fVSRN & single & \textcolor{black}{29.110} & \textcolor{black}{55s} \\
 &  & NGP & single & 28.391 & 1m 33s \\
 &  & APMGSRN & single & {\ul \textbf{29.424}} & 51s \\ \cline{2-6} 
 & \multirow{3}{*}{Medium - 4 MB} & fVSRN & single & \textcolor{black}{37.022} & \textcolor{black}{57s} \\
 &  & NGP & single & 35.142 & 1m 8s \\
 &  & APMGSRN & single & {\ul \textbf{37.807}} & 1m 34s \\ \cline{2-6} 
 & \multirow{4}{*}{Large - 64 MB} & fVSRN & single & \textcolor{black}{42.092} & \textcolor{black}{1m 46s} \\
 &  & NGP & single & 42.750 & 2m 10s \\
 &  & APMGSRN & single & 44.075 & 3m 7s \\
 &  & APMGSRN & $2 \times 2 \times 2$ & {\ul \textbf{46.118}} & 1m 35s \\ \hline
\multirow{10}{*}{\begin{tabular}[c]{@{}c@{}}Supernova\\ $432^3$\\ 403 MB\\ Entropy in supernova\\ simulation. \\ Somewhat sparse.\end{tabular}} & \multirow{3}{*}{Small - 280 KB} & fVSRN & single & \textcolor{black}{39.505} & \textcolor{black}{54s} \\
 &  & NGP & single & 41.523 & 1m 29s \\
 &  & APMGSRN & single & {\ul \textbf{41.891}} & 1m 16s \\ \cline{2-6} 
 & \multirow{3}{*}{Medium - 4 MB} & fVSRN & single & \textcolor{black}{42.697} & \textcolor{black}{56s} \\
 &  & NGP & single & 44.811 & 1m 8s \\
 &  & APMGSRN & single & {\ul \textbf{46.831}} & 1m 19s \\ \cline{2-6} 
 & \multirow{4}{*}{Large - 64 MB} & fVSRN & single & \textcolor{black}{46.271} & \textcolor{black}{1m 46s} \\
 &  & NGP & single & 47.806 & 2m 32s \\
 &  & APMGSRN & single & 49.880 & 2m 41s \\
 &  & APMGSRN & $2 \times 2 \times 2$ & {\ul \textbf{50.644}} & 35s \\ \hline
\multirow{10}{*}{\begin{tabular}[c]{@{}c@{}}Asteroid\\ $1000^3$\\ 3.73 GB\\ Volume fraction of asteroid \\ plus water fields from \\ simulation of an \\ asteroid hitting the ocean. \\ SciVis contest 2018 data. \\ Very sparse.\end{tabular}} & \multirow{3}{*}{Small - 280 KB} & fVSRN & single & \textcolor{black}{31.072} & \textcolor{black}{54s} \\
 &  & NGP & single & {\ul \textbf{33.713}} & 1m 32s \\
 &  & APMGSRN & single & 30.130 & 1m 43s \\ \cline{2-6} 
 & \multirow{3}{*}{Medium - 4 MB} & fVSRN & single & \textcolor{black}{33.783} & \textcolor{black}{54s} \\
 &  & NGP & single & {\ul \textbf{39.012}} & 1m 6s \\
 &  & APMGSRN & single & 35.799 & 1m 49s \\ \cline{2-6} 
 & \multirow{4}{*}{Large - 64 MB} & fVSRN & single & \textcolor{black}{39.121} & \textcolor{black}{1m 47s} \\
 &  & NGP & single & 42.928 & 2m 24s \\
 &  & APMGSRN & single & 39.455 & 2m 49s \\
 &  & APMGSRN & $4 \times 4 \times 4$ & {\ul \textbf{43.188}} & 45s \\ \hline
\multirow{10}{*}{\begin{tabular}[c]{@{}c@{}}Isotropic \cite{JHUTDB1,JHUTDB2}\\ $1024^3$\\ 4 GB\\ Velocity magnitude field \\ of forced isotropic \\ turbulence. \\ Not sparse.\end{tabular}} & \multirow{3}{*}{Small - 280 KB} & fVSRN & single & \textcolor{black}{27.379} & \textcolor{black}{58s} \\
 &  & NGP & single & 27.007 & 1m 32s \\
 &  & APMGSRN & single & {\ul \textbf{27.622}} & 55s \\ \cline{2-6} 
 & \multirow{3}{*}{Medium - 4 MB} & fVSRN & single & \textcolor{black}{31.570} & \textcolor{black}{58s} \\
 &  & NGP & single & 30.786 & 1m 8s \\
 &  & APMGSRN & single & {\ul \textbf{31.825}} & 1m 14s \\ \cline{2-6} 
 & \multirow{4}{*}{Large - 64 MB} & fVSRN & single & \textcolor{black}{36.858} & \textcolor{black}{1m 46s} \\
 &  & NGP & single & 37.198 & 2m 10s \\
 &  & APMGSRN & single & 38.120 & 3m 23s \\
 &  & APMGSRN & $4 \times 4 \times 4$ & {\ul \textbf{41.086}} & 57s \\ \hline
\multirow{7}{*}{\begin{tabular}[c]{@{}c@{}}Rotstrat \cite{JHUTDB1, JHUTDB2}\\ $4096^3$\\ 250 GB\\ Velocity magnitude of \\ rotating stratified \\ turbulence, t=4. \\ Not sparse.\end{tabular}} & \multirow{7}{*}{\begin{tabular}[c]{@{}c@{}}Small - 55 MB\\ Medium - 183 MB\\ Large - 864 MB\end{tabular}} & \multirow{7}{*}{\begin{tabular}[c]{@{}l@{}}APMGSRN\\ APMGSRN\\ APMGSRN\end{tabular}} & \multirow{7}{*}{\begin{tabular}[c]{@{}l@{}}$3 \times 3 \times 3$\\ $3 \times 3 \times 3$\\ $3 \times 3 \times 3$\end{tabular}} & \multirow{7}{*}{\begin{tabular}[c]{@{}l@{}}41.322\\ 44.294\\ 49.074\end{tabular}} &  \\
 &  &  &  &  &  \\
 &  &  &  &  & 1m 3s \\
 &  &  &  &  & 1m 15s \\
 &  &  &  &  & 1m 58s \\
 &  &  &  &  &  \\
 &  &  &  &  &  \\ \hline
\begin{tabular}[c]{@{}c@{}}Channel \cite{channel, JHUTDB1, JHUTDB2}\\ $7680 \times 1568 \times 10240$\\ 450GB\\ Channel flow at Reynolds \\ number $\approx 5200$, t=10. \\ Not sparse.\end{tabular} & 97 MB & APMGSRN & $4 \times 1 \times 6$ & 40.055 & 35s \\
 & \multicolumn{1}{l|}{} &  &  &  &  \\
 & \multicolumn{1}{l|}{} &  &  &  & 
\end{tabular}
}
\vspace{-5mm}
\end{table}

\section{Experiments}
\label{experiments}

We perform experiments to test the training and reconstruction metrics of our proposed APMGSRN model in \autoref{SRNcomparison}.
Our domain decomposition training approach is evaluated in \autoref{domaindecompositionevaluation}.
Lastly, we visualize trained models with our renderer in \autoref{rendererevaluation}.
Our datasets and brief descriptions are shown in \autoref{quanttable}.

\textbf{Hyperparameters and evaluation hardware}.
To save page space, model hyperparameters for each size SRN (small/medium/large/\textcolor{black}{decomposition}) are available in GitHub in the ``BatchRunSettings'' folder, as there are many hyperparameters to enumerate for each architecture.
For our model specifically, the number of grids we use is either 16 or 32, the grid resolution varies from $16^3$ to $64^3$, and the number of features per feature grid vertex is either 1 or 2.
\textcolor{black}{We find that inference speed degrades above 64 grids, and accuracy does not improve significantly with more grids or more features per grid to justify the increased storage cost.}
Grid resolution has the largest impact on both network performance and network storage size.

Training for all models and compression tests are run on a remote NVidia DGX system with 8 NVidia 40GB A100 GPUs, a 64-core AMD EPYC 7742 processor, and 1TB of system memory connected to a Lustre-based file system.
Rendering evaluation is performed on a Windows 11 PC with a NVidia 2080 Ti, a 12-core AMD Ryzen 9 3900X, and 32GB of system memory.

\textbf{Baseline SRNs.}
Neural graphics primitives (NGP) \cite{Muller22_ngp} uses a hierarchical hash grid encoding scheme that uses a random hash function to map spatial coordinates to hash table entries which hold feature values.
The encoding scheme allows hash collisions and copes with them by letting the gradients average, which puts more emphasis on the higher-error regions.
Therefore, NGP benefits from very high sparsity, where hash collisions are frequent, but a majority of the space is empty anyway.
The second model we compare with is fast volumetric scene representation network (fVSRN) \cite{Weiss22_fvsrn}, which is a standard feature-grid based SRN composed with Fourier frequency encoding \cite{tancik20_fourier}.
The feature grid is dense and lines up with the extents of the data.
For both comparison models, we use initialization and training routines as suggested by the publication they were introduced in. 

\textbf{Training.}
Python 3.9 with PyTorch \cite{NEURIPS19_pytorch} are used for training all networks.
All models use the same fully-connected decoder from Tiny-CUDA-NN (TCNN) \cite{muller21_tcnn} with 2 layers of 64 neurons, accelerated with custom fully-fused CUDA code.
\textcolor{black}{NGP's hash grid encoding \cite{muller21_tcnn} native implementation from TCNN is used, and we implement fVSRN using TCNN for high efficiency, but our own model uses PyTorch with single precision for the multi-grid encoding.}
All networks are trained for 50,000 iterations or until an early stopping criteria, as discussed in \autoref{trainingroutine}.
For all models, the Adam \cite{adam} optimizer is used with $\beta_1=0.9, \beta_2=0.99$ with a learning rate of 0.01.
Our transformation matrices use a learning rate of 0.001.
Each iteration, a batch of 100,000 coordinates are sampled uniformly from $[-1, 1]^3$, and the ground truth values are trilinearly interpolated from the volume.

\subsection{Scene representation network comparison}
\label{SRNcomparison}

We compare our model against two state-of-the-art scene representation networks, neural graphics primitives (NGP) \cite{Muller22_ngp} and fVSRN \cite{Weiss22_fvsrn}, for data reconstruction quality and training times at fixed model sizes.

\textbf{Single network performance.}
Across results from each dataset and network size shown in \autoref{quanttable}, APMGSRN is first for reconstruction peak signal-to-noise ration (PSNR - higher is better) in all but \textcolor{black}{3 settings}.
APMGSRN's adaptivity seems most dominant in somewhat sparse data, such as plume, Nyx, and supernova, where it outperforms other models by 2+ dB at the largest model size.
Our model underperforms against NGP only for the asteroid dataset because of the high degree of sparsity, which NGP excels at. 
Examples of learned grid positions are shown in \autoref{fig:teaser}, and verifies that the grids learn to migrate to regions with complex features regardless of dataset.

Training times across models are similar, with our model being the least performant compared to NGP and fVSRN. 
This is a consequence of using a second loss function for updating our transformation matrices, and more matrix operations per model forward pass due to the transformations required before feature grid interpolation.
Our multi-grid encoding is a pure PyTorch implementation, but custom CUDA code like that used by NGP, could improve model performance.

\begin{figure}[ht]
\centering
  \includegraphics[width=0.8\columnwidth]{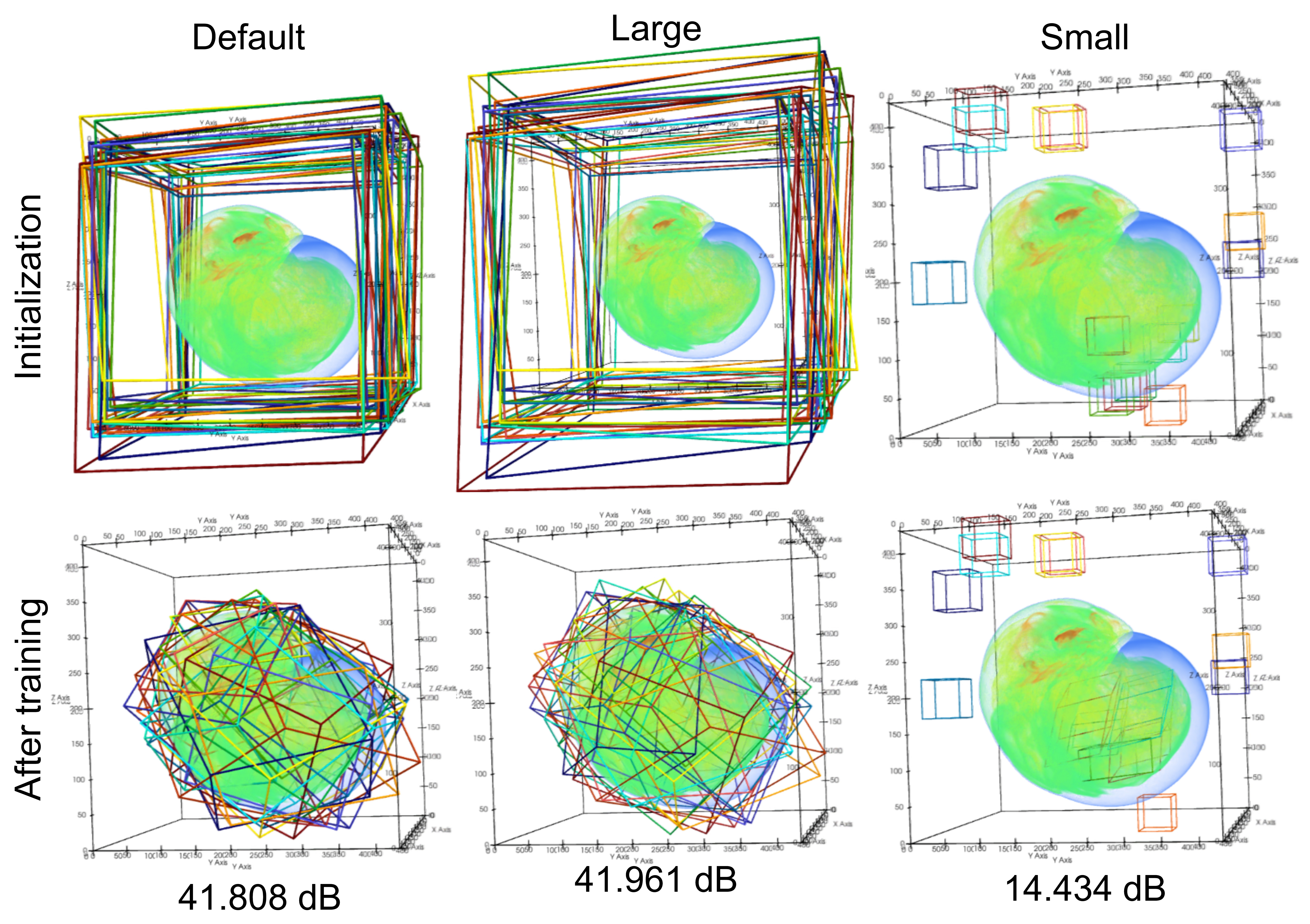}
  \caption{
    Examples of 3 grid initializations and the effect on final grid locations and data reconstruction PSNR. 
    Underlying data volume rendered depicts ground truth data for clarity of grid locations with respect to the raw data.
  }
  \label{fig:initialization}
\end{figure}

\textbf{Importance of grid initialization and grid visualization.}
As feature grid location in the volume is the only means of encoding coordinates to a high-dimensional feature space for learning, the initialization of the grids may affect learning quality.
We experiment with grid initialization with a small sized model trained on the supernova dataset. 
We use three initialization techniques for the grids: (1) our recommended initialization as described in \autoref{initialization}, (2) our recommended initialization, with the grid scales increased by 20\% so as to cover extra empty space outside of the volume, (3) initializing grids at a small scale, randomly distributed within the domain.
The results of the experiment are shown in \autoref{fig:initialization}.
We can see that the final grids and reconstruction quality of the first two initialization schemes are similar, but initializing the grids to random spots with a small scale performed much worse.
The grids that do not cover any data sitting in empty space have near-zero gradient for changing any of the transformation matrix parameters.
A larger learning rate or a less powerful flat-top gaussian can make up for this for some cases, but that may cause larger-scale grids to see unstable updates to their parameters.
For best results, we recommend our default initialization of global-scale grids with small random perturbations, but we believe there may be room for improved performance with different differentiable approximations of grid density for smoother learning.
We recommend readers view our supplemental material for videos of feature grids during training on the smaller datasets experimented with, as it gives a strong intuition for what the network is learning and how it is learning it.
\textcolor{black}{In the future, other grid initialization strategies may prove useful for better learned grid positions.
Additionally, we believe that scaling up the flat-top gaussian strength (starting from 1.0) during training can help situations like our small initialization scheme, because gradients will not approach 0 rapidly for data outside the grid.}

\begin{figure}[ht]
\centering
  \includegraphics[width=1.0\columnwidth]{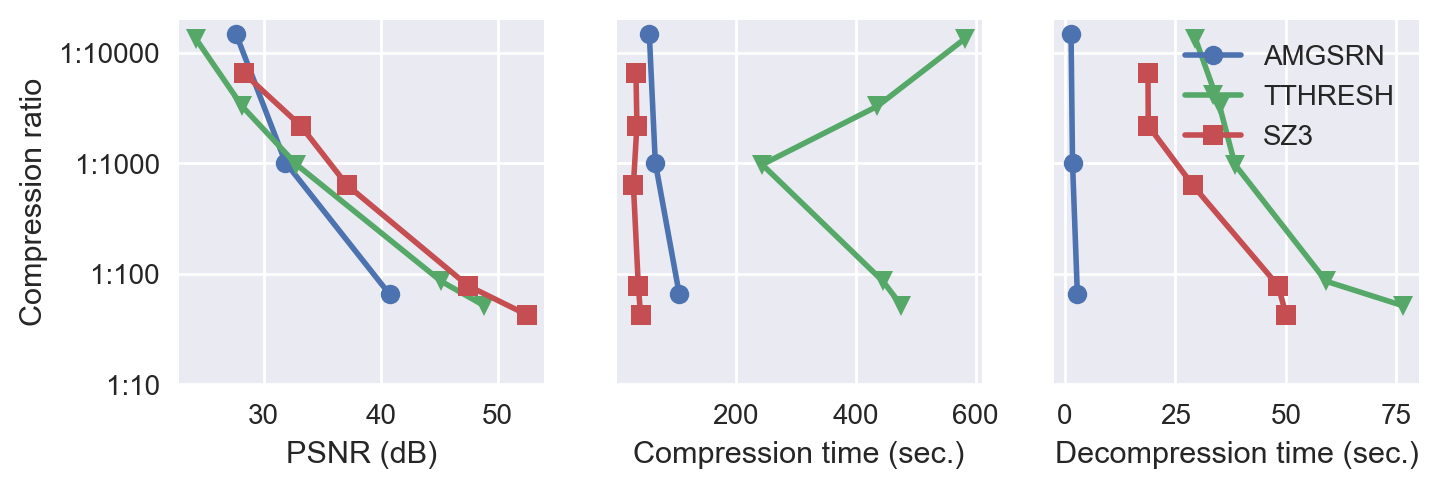}
  \caption{
    Comparison of APMGSRN with TTHRESH and SZ for compressing the isotropic volume ($1024^3$). 
    Our ``compression'' time is the time it takes to train the model, and our decompression time is the time it takes to query each point in the volume.
  }
  \label{fig:compression}
\end{figure}

\textbf{Compression.}
Though data reduction is not the only intended use case for our approach, nor is it something we design for specifically (such as with network weight quantization for further data reduction \cite{Lu21_neurocomp}), we believe it is useful to the community to compare the compression ability of state-of-the-art SRNs with state-of-the-art compressors.
We compare compression results from TTHRESH \cite{Ballester20_tthresh} and SZ3 \cite{zhao21_sz3_3, liang18_sz3_2, liang23_sz3_1} with our approach in \autoref{fig:compression}.
We do not compare to a rendering-focused compressor such as cudaCompress \cite{treib12_cudacompress} or a bricked version of TTHRESH \cite{Ballester20_tthresh}, as this has been done by recent work \cite{Weiss22_fvsrn} showing that its decoding is significantly slower,  compression rates smaller, memory use is higher, and image quality is worse than state-of-the-art SNRs.

As shown in \autoref{fig:compression}, our approach only provides a higher compression ratio than TTHRESH or SZ3 at a $10000\times$ data reduction rate with a reconstruction quality of under 30 dB PSNR, and is otherwise less compressive than state-of-the-art compressors.
The throughput (decompression) from APMGSRN is much quicker regardless of compression level, decoding on the order of 500 million points per second, but keep in mind that APMGSRN is decoding on the GPU whereas SZ3 and TTHRESH are on CPU.
The largest difference between APMGSRN and these compressors is that our approach can efficiently perform arbitrary point evaluations, whereas SZ3 and TTHRESH both require decompressing to the original data size before trilinear interpolation.

We also test compression on our two large datasets, rotstrat and channel. 
The tested compressors take significantly longer to compress our large-scale data, or fail to do so at all. 
In fact, TTHRESH and SZ3 both run out of memory on our machine with 1TB of memory when trying to compress the channel dataset (450GB), and TTHRESH also runs out of memory when trying to compress the rotstrat dataset (250GB).
SZ3 successfully compresses the 250GB rotstrat data in 38 minutes, resulting in a compression ratio of $509 \times$ and a reconstruction quality of 20.60 dB PSNR after another 14 minutes of decompression.
Both our small and medium sized (domain decomposition) models for rotstrat (see \autoref{quanttable}) achieve higher PSNRs (41.32 dB and 44.29 dB) with higher compression rates ($4791\times$ and $1432\times$) while only needing to train for 34 minutes if all models are trained sequentially on 1 GPU, or 5 minutes in parallel on 8 GPUs.
Additionally, our decoding takes 3 minutes for the whole $4096^3$ volume.

\subsection{domain decomposition network evaluation}
\label{domaindecompositionevaluation}

We evaluate our domain decomposition training approach on each dataset, and show results in \autoref{quanttable}.
Even though the smaller datasets do not require \textcolor{black}{the domain decomposition} to train without memory limitations, the domain decomposition model outperforms a single model at the same storage size for each dataset.
This is expected, as each of the domain decomposition models will have their own decoder and set of feature grids that may create and advantage over the single large model.
The downside of a domain decomposition model is that training may take longer if you only have 1 GPU and train serially.

For data that cannot fit in a single GPU, such as rotstrat and channel, our domain decomposition approach provides an efficient alternative to an out-of-core training routine.
In \autoref{quanttable}, we list the \textit{average} training time per model for the domain decomposition models. 
Our training machine had 8 GPUs training in parallel and the total training time was \textcolor{black}{7 minutes with data I/O for rotstrat.}

\begin{figure}[ht]
\centering
  \includegraphics[width=0.9\columnwidth]{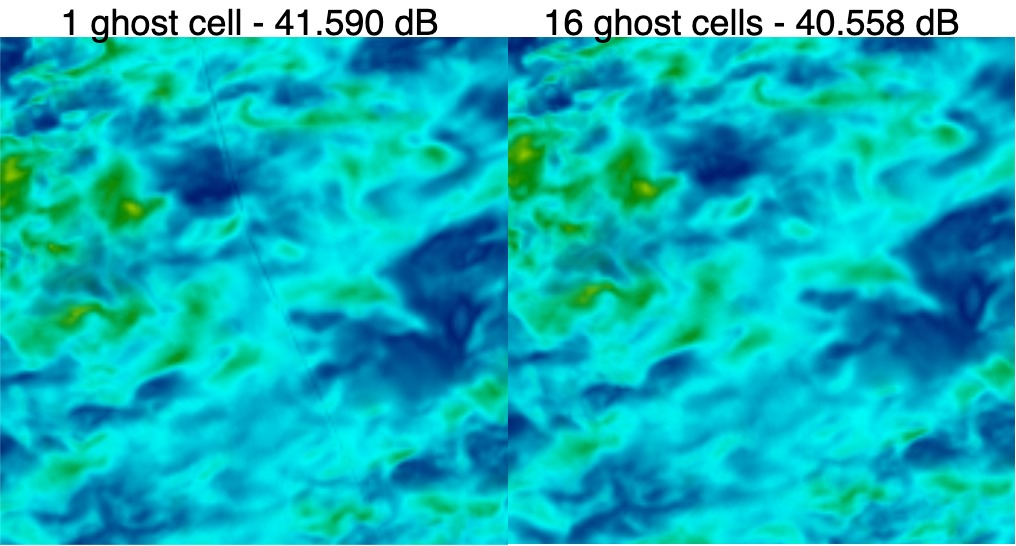}
  \caption{
   Zoom in of volume render from identical $4 \times 4 \times 4$ domain decomposition networks trained with either \textcolor{black}{1} or 16 ghost cells. A seam is clear with 1 ghost cell.
  }
  \label{fig:ghostcells}
\end{figure}

\textbf{Effect of ghost cells.}
By default, our domain decomposition has 1 ghost cell for corner alignment across models.
We show the effect of increasing ghost cells within our domain decomposition models in \autoref{fig:ghostcells}, which is a volume rendering of \textcolor{black}{two networks - one trained with the default of 1 ghost cell, and one with 16 ghost cells.}
The boundary artifact is visible only in \textcolor{black}{the render with only 1 ghost cell}.
However, a significant PSNR drop of -1 dB is noticed with this increase in ghost cells, attributed to the fact that a network \textcolor{black}{in the decomposition} learning a volume of size $288^3$ is a 42\% larger volume than the original $256^3$ volume learned on in the version without ghost cells.
The redundancy of learning a significant portion of the full volume in multiple networks reduces the learning capacity of the \textcolor{black}{set of networks}.
Since the artifacts are fairly minor while reducing performance, we recommend using 1 or few ghost cells for general volume rendering, but 16 ghost cells for tasks like isosurface extraction where seams can be very distracting.
We consider future work for reducing the boundary effect between networks with a network communication scheme during training.

\begin{figure}[ht]
\centering
  \includegraphics[width=0.8\columnwidth]{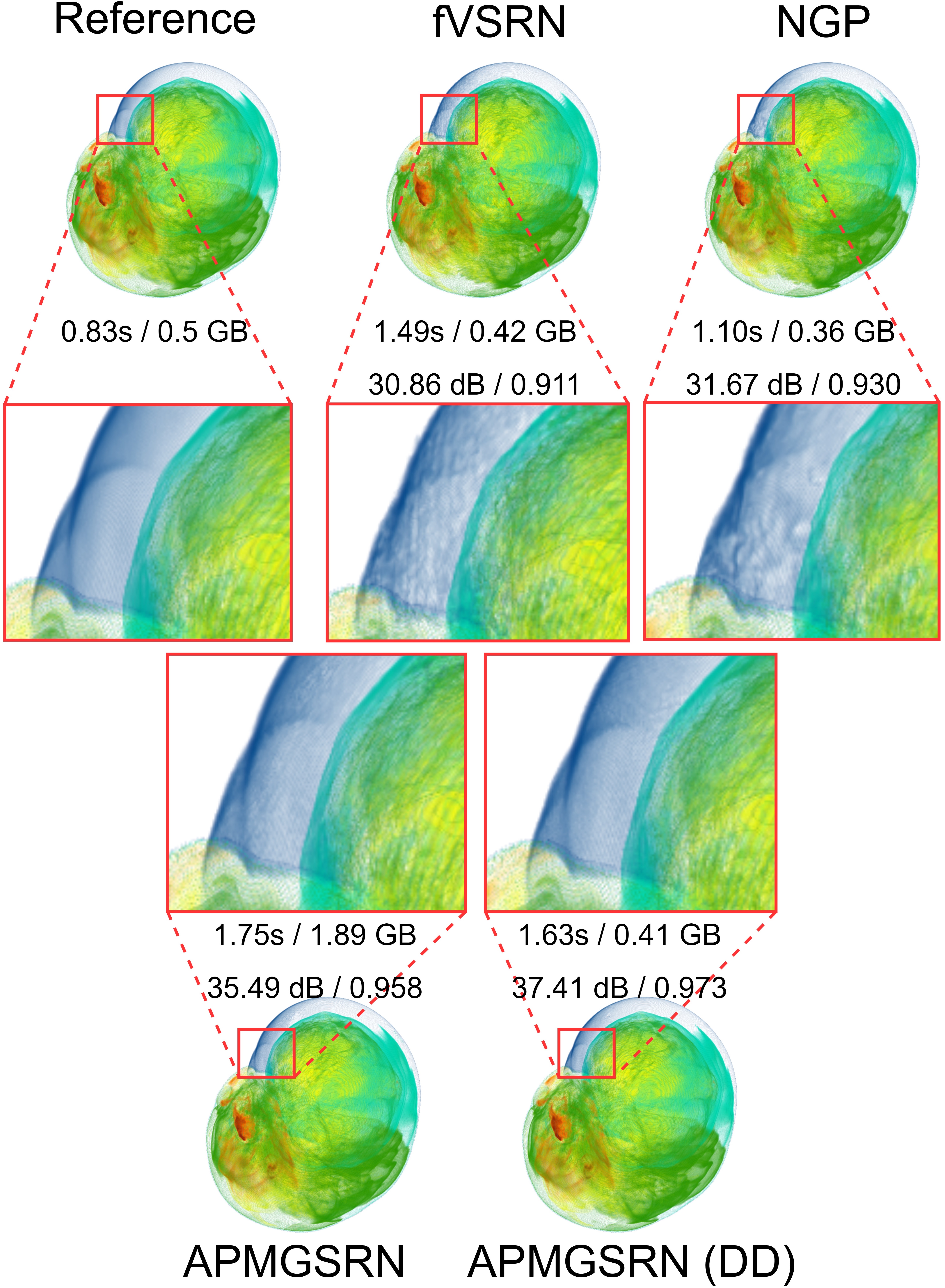}
  \caption{\textcolor{black}{
    $1024^2$ volume renders of Supernova data ($432^3$) with a step size of 1 voxel. A batch size of $2^{23}$ is used during forward passes. The reference is the raw data (308 MB), while the others are neural representations of the data in their ``large'' configuration (64 MB), where (DD) means domain decomposition. The first row of metrics listed is render time / memory use, and the 2nd row is PSNR (dB) / SSIM. Metrics are each for the entire image, not the zoom in.}
  }
  \label{fig:qualitative}
\end{figure}

\begin{figure}[ht]
\centering
  \includegraphics[width=\columnwidth]{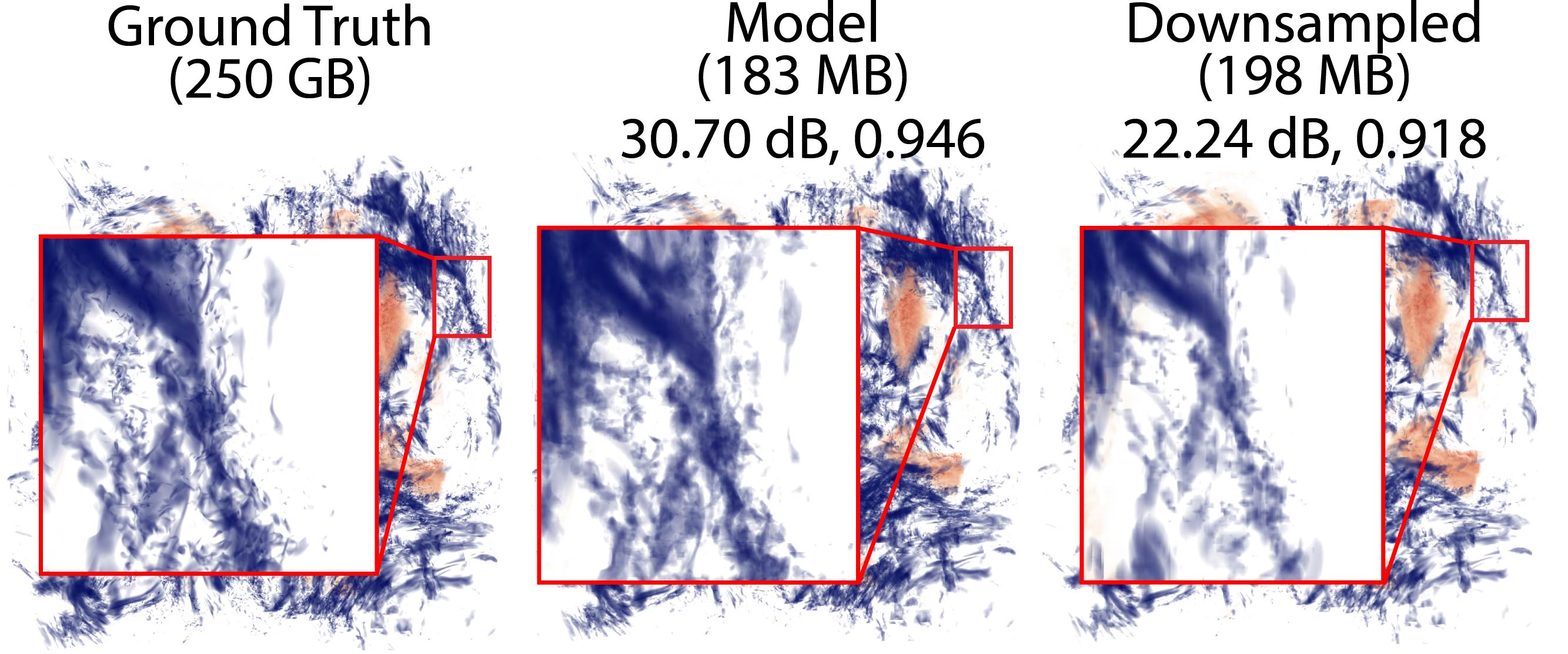}
  \caption{
    Volume renders of the original rotstrat data, our ``medium'' size model representing the data, and a low resolution version of the original data, which has been subsampled by 11 in each direction to have a similar storage footprint as the model.
    Each image is rendered at $8000^2$ with the same visual mapping parameters.
    Metrics listed are PSNR/SSIM for the entire image (not just zoomed portion).
  }
  \label{fig:downsampled}
\end{figure}

\begin{figure}[ht]
\centering
  \includegraphics[width=\columnwidth]{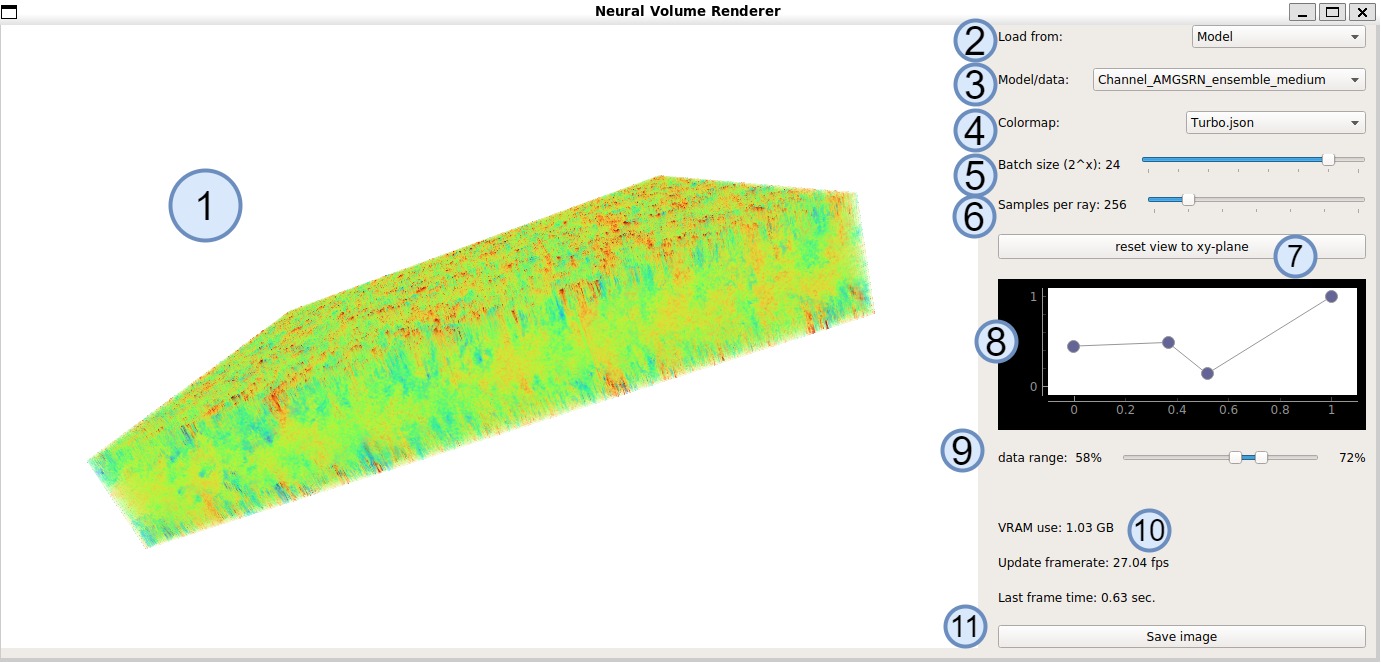}
  \caption{
    The interface for our volume renderer with a volume render of an domain decomposition model trained on the 450GB channel dataset. \textbf{(1)} The viewing area for interaction. \textbf{(2)} Dropdown to choose to load a model or data (NetCDF). \textbf{(3)} Which model/data to load from. \textbf{(4)} Colormap to use. \textbf{(5)} Batch size when querying the network. Larger reduces total frame time, but requires more GPU memory and slows down interactivity. \textbf{(6)} Samples per ray. \textbf{(7)} Resets camera view on data. \textbf{(8)} Opacity transfer function editor. \textbf{(9)} Relative min/max for transfer function. \textbf{(10)} Render statistics. \textbf{(11)} Save current frame button.
  }
  \label{fig:renderer}
\end{figure}

\subsection{Neural volume rendering}
\label{rendererevaluation}

In this section, we evaluate the single frame performance of neural volume rendering.
The raymarching and compositing are done with the same code, so the timing and memory use difference is accounted for solely by querying the SRN or raw volume data.

Our single frame rendering results are shown in \autoref{fig:qualitative} where we compare the render results of the large sized (64 MB on disk) models for the supernova dataset.
The reference render using the volume is the quickest, but assumes the volume can fit in GPU memory, which is not possible for datasets larger than 11 GB on our test machine.
NGP provides the quickest render times of the networks on our 2080Ti, as well as the lowest memory use.
\textcolor{black}{fVSRN is slower likely due to the Fourier frequency encoding.}
\textcolor{black}{Our models are slowest to render,} but have considerably better image reconstruction shown by the higher PSNR and structural similarity index measure (SSIM) \cite{SSIM}.
Our approach (single model) requires the most memory due to the transformation to the 64 local grid spaces, requiring $64\times$ the memory of the global coordinates for the local space coordinates.

In \autoref{fig:downsampled}, we compare our trained ``medium'' sized domain decomposition model for the 250 GB rotstrat data with a downsampled version of the raw data with a similar storage footprint as the model.
We subsample the raw data by 11 in each dimension to create a volume that takes 198 MB to store, which is only slightly larger than our network which takes 183 MB to store. 
We use our renderer to generate $8000^2$ images of each representation of the data with the same visual mapping parameters.
The image from our model achieves a significantly higher 30.70 dB PSNR compared with the render from the downsampled data which only achieves 22.24 dB.
Additionally, the neural network captures features that are entirely missed by the subsampled data.

We include an image of our rendering application in \autoref{fig:renderer}, but do not discuss it in detail here.
Please see supplemental materials for a video using the tool in real-time\textcolor{black}{, as well as additional qualitative comparisons of volume renders from each model}.

\section{Limitations}
\label{limitations}

One limitation of our model is that memory use blows up and training speeds reduce as the number of feature grids increases.
This is from the global to local transformation, which needs to allocate memory equal to the memory of the query coordinates times the number of feature grids, and then perform interpolation for each of those coordinates.
Further engineering effort could reduce the overhead with custom CUDA code or by limiting the number of grids operated on at once.

Though our neural renderer is compatible with any PyTorch model coded in Python, the renderer is slower than what is reported by contemporary work \textcolor{black}{\cite{Wu22_ngpscivis, Weiss22_fvsrn, wu2023_hyperinr}}.
\textcolor{black}{Further optimizations to the code could improve the rendering speed while remaining flexible for new PyTorch models to be plugged in with ease.}

\section{Conclusion and future work}
\label{conclusion}
In this work, we present a novel SRN called APMGSRN, a model which excels at representing scientific data due to the adaptivity of the model.
In order to train on datasets that cannot fit in-core during training, we propose an domain decomposition approach for training and inference that divides the domain into separate chunks to be learned by multiple networks.
In our evaluations, we find that our proposed APMGSRN architecture outperforms the reconstruction performance of other state-of-the-art models at similar model sizes in both data space and image space.
Our model cannot compress as well as state-of-the-art compressors like TTHRESH, but achieves high compression rate with efficient volume rendering in the compression domain without \textcolor{black}{the need} to decompress to the original data's size.

In the future, we think our adaptive feature grids can be augmented in two ways: (1) the feature grids need not be trilinearly interpolated - another encoding scheme, such as a hash grid from NGP, can be used for query within each grid, (2) the feature grids could extend the time dimension, creating 4D learnable feature grids supporting temporal interpolation.
Reducing boundary artifacts is another key direction to improve upon to reduce distracting seams in visualizations.
On the engineering side, there are many places where optimizations can improve efficiency, such as the multi-grid encoder or domain decomposition model inference.
Our approach could also work with minimal changes to represent non-uniform grid data types, such as unstructured, point-based, or AMR, so long as the data can return a scalar value for an arbitrary point query.
With the increasing popularity and usefulness of SRNs, future work to support neural volume rendering in mature open-source software such as ParaView would greatly benefit the visualization community.


\acknowledgments{%
This work is supported in part by the US Department of Energy SciDAC program DE-SC0021360, National Science Foundation Division of Information and Intelligent Systems IIS-1955764, and National Science Foundation Office of Advanced Cyberinfrastructure OAC-2112606 and OAC-2311878. 
This work is also supported by Advanced Scientific Computing Research, Office of Science, U.S. Department of Energy, under Contracts DE-AC02-06CH11357, DE-SC0022753, DE-SC0023193 under program manager Margaret Lentz.%
}

\bibliographystyle{abbrv-doi-hyperref}

\bibliography{template}

\begin{thebibliography}{10}

\bibitem{nyx}
A.~S. Almgren, J.~B. Bell, M.~J. Lijewski, Z.~Luki{\'{c}}, and E.~V. Andel.
\newblock {Nyx: A Massively Parallel AMR Code for Computational Cosmology}.
\newblock {\em The Astrophysical Journal}, 765(1):39--53, 2013.

\bibitem{Ballester20_tthresh}
R.~Ballester-Ripoll, P.~Lindstrom, and R.~Pajarola.
\newblock {TTHRESH: Tensor Compression for Multidimensional Visual Data}.
\newblock {\em IEEE Transactions on Visualization and Computer Graphics},
  26(9):2891--2903, 2020. \href{https://doi.org/10.1109/TVCG.2019.2904063}
{doi: {{%
10\hspace{.1pt}\discretionary{.}{%
}{.}\hspace{.4pt}1109\discretionary{/}{%
}{/}TVCG\hspace{.1pt}\discretionary{.}{%
}{.}\hspace{.4pt}2019\hspace{.1pt}\discretionary{.}{%
}{.}\hspace{.4pt}2904063}}}


\bibitem{Chen22_tensorf}
A.~Chen, Z.~Xu, A.~Geiger, J.~Yu, and H.~Su.
\newblock {TensoRF: Tensorial Radiance Fields}.
\newblock In {\em Proc. European Conference on Computer Vision}, pp. 333--350,
  2022.

\bibitem{genova2020_local}
K.~Genova, F.~Cole, A.~Sud, A.~Sarna, and T.~Funkhouser.
\newblock Local deep implicit functions for 3d shape.
\newblock In {\em Proceedings of the IEEE/CVF Conference on Computer Vision and
  Pattern Recognition}, pp. 4857--4866, 2020.

\bibitem{glorot10_xavierinit}
X.~Glorot and Y.~Bengio.
\newblock Understanding the difficulty of training deep feedforward neural
  networks.
\newblock In {\em Proceedings of the Thirteenth International Conference on
  Artificial Intelligence and Statistics}, vol.~9, pp. 249--256, 13--15 May
  2010.

\bibitem{han22_coordnet}
J.~Han and C.~Wang.
\newblock {CoordNet: Data Generation and Visualization Generation for
  Time-Varying Volumes via a Coordinate-Based Neural Network}.
\newblock {\em IEEE Transactions on Visualization and Computer Graphics}, pp.
  1--12, 2022. \href{https://doi.org/10.1109/TVCG.2022.3197203}
{doi: {{%
10\hspace{.1pt}\discretionary{.}{%
}{.}\hspace{.4pt}1109\discretionary{/}{%
}{/}TVCG\hspace{.1pt}\discretionary{.}{%
}{.}\hspace{.4pt}2022\hspace{.1pt}\discretionary{.}{%
}{.}\hspace{.4pt}3197203}}}


\bibitem{Höhlein22_vrncompression}
K.~Höhlein, S.~Weiss, and R.~Westermann.
\newblock {Evaluation of Volume Representation Networks for Meteorological
  Ensemble Compression}.
\newblock In {\em Vision, Modeling, and Visualization}. The Eurographics
  Association, 2022. \href{https://doi.org/10.2312/vmv.20221198}
{doi: {{%
10\hspace{.1pt}\discretionary{.}{%
}{.}\hspace{.4pt}2312\discretionary{/}{%
}{/}vmv\hspace{.1pt}\discretionary{.}{%
}{.}\hspace{.4pt}20221198}}}


\bibitem{adam}
D.~P. Kingma and J.~Ba.
\newblock {Adam: {A} Method for Stochastic Optimization}.
\newblock In {\em Proc. 2015 International Conference on Learning
  Representations}, 2015.

\bibitem{channel}
M.~Lee and R.~D. Moser.
\newblock {Direct numerical simulation of turbulent channel flow up to
  Re$_{\tau}$ = 5200}.
\newblock {\em Journal of Fluid Mechanics}, 774:395--415, 2015.

\bibitem{Li22_nerfacc}
R.~Li, M.~Tancik, and A.~Kanazawa.
\newblock {NerfAcc: A General NeRF Acceleration Toolbox}, 2022.
  \href{https://doi.org/10.48550/ARXIV.2210.04847}
{doi: {{%
10\hspace{.1pt}\discretionary{.}{%
}{.}\hspace{.4pt}48550\discretionary{/}{%
}{/}ARXIV\hspace{.1pt}\discretionary{.}{%
}{.}\hspace{.4pt}2210\hspace{.1pt}\discretionary{.}{%
}{.}\hspace{.4pt}04847}}}


\bibitem{JHUTDB1}
Y.~Li, E.~Perlman, M.~Wan, Y.~Yang, C.~Meneveau, R.~Burns, S.~Chen, A.~Szalay,
  and G.~Eyink.
\newblock {A public turbulence database cluster and applications to study
  Lagrangian evolution of velocity increments in turbulence}.
\newblock {\em Journal of Turbulence}, 9, 2008.

\bibitem{liang18_sz3_2}
X.~Liang, S.~Di, D.~Tao, S.~Li, S.~Li, H.~Guo, Z.~Chen, and F.~Cappello.
\newblock Error-controlled lossy compression optimized for high compression
  ratios of scientific datasets.
\newblock In {\em 2018 IEEE International Conference on Big Data (Big Data)},
  pp. 438--447, 2018. \href{https://doi.org/10.1109/BigData.2018.8622520}
{doi: {{%
10\hspace{.1pt}\discretionary{.}{%
}{.}\hspace{.4pt}1109\discretionary{/}{%
}{/}BigData\hspace{.1pt}\discretionary{.}{%
}{.}\hspace{.4pt}2018\hspace{.1pt}\discretionary{.}{%
}{.}\hspace{.4pt}8622520}}}


\bibitem{liang23_sz3_1}
X.~Liang, K.~Zhao, S.~Di, S.~Li, R.~Underwood, A.~M. Gok, J.~Tian, J.~Deng,
  J.~C. Calhoun, D.~Tao, Z.~Chen, and F.~Cappello.
\newblock {SZ3: A Modular Framework for Composing Prediction-Based
  Error-Bounded Lossy Compressors}.
\newblock {\em IEEE Transactions on Big Data}, 9(2):485--498, 2023.
  \href{https://doi.org/10.1109/TBDATA.2022.3201176}
{doi: {{%
10\hspace{.1pt}\discretionary{.}{%
}{.}\hspace{.4pt}1109\discretionary{/}{%
}{/}TBDATA\hspace{.1pt}\discretionary{.}{%
}{.}\hspace{.4pt}2022\hspace{.1pt}\discretionary{.}{%
}{.}\hspace{.4pt}3201176}}}


\bibitem{lindell21_autoint}
D.~B. Lindell, J.~N.~P. Martel, and G.~Wetzstein.
\newblock {AutoInt: Automatic Integration for Fast Neural Volume Rendering}.
\newblock In {\em Proc. CVPR}, 2021.

\bibitem{liu20_nsvf}
L.~Liu, J.~Gu, K.~Z. Lin, T.-S. Chua, and C.~Theobalt.
\newblock Neural sparse voxel fields.
\newblock {\em NeurIPS}, 2020.

\bibitem{Lu21_neurocomp}
Y.~Lu, K.~Jiang, J.~A. Levine, and M.~Berger.
\newblock Compressive neural representations of volumetric scalar fields.
\newblock {\em Computer Graphics Forum}, 40(3):135--146, 2021.
  \href{https://doi.org/10.1111/cgf.14295}
{doi: {{%
10\hspace{.1pt}\discretionary{.}{%
}{.}\hspace{.4pt}1111\discretionary{/}{%
}{/}cgf\hspace{.1pt}\discretionary{.}{%
}{.}\hspace{.4pt}14295}}}


\bibitem{Martel21_acorn}
J.~N.~P. Martel, D.~B. Lindell, C.~Z. Lin, E.~R. Chan, M.~Monteiro, and
  G.~Wetzstein.
\newblock {Acorn: Adaptive Coordinate Networks for Neural Scene
  Representation}.
\newblock {\em ACM Trans. Graph.}, 40(4), 2021.
  \href{https://doi.org/10.1145/3450626.3459785}
{doi: {{%
10\hspace{.1pt}\discretionary{.}{%
}{.}\hspace{.4pt}1145\discretionary{/}{%
}{/}3450626\hspace{.1pt}\discretionary{.}{%
}{.}\hspace{.4pt}3459785}}}


\bibitem{mildenhall20_nerf}
B.~Mildenhall, P.~P. Srinivasan, M.~Tancik, J.~T. Barron, R.~Ramamoorthi, and
  R.~Ng.
\newblock {NeRF: Representing Scenes as Neural Radiance Fields for View
  Synthesis}.
\newblock In {\em Proc. European Conference on Computer Vision}, 2020.

\bibitem{muller21_tcnn}
T.~M\"uller.
\newblock {tiny-cuda-nn}.
\newblock \url{https://github.com/NVlabs/tiny-cuda-nn}, 4 2021.

\bibitem{Muller22_ngp}
T.~M\"{u}ller, A.~Evans, C.~Schied, and A.~Keller.
\newblock Instant neural graphics primitives with a multiresolution hash
  encoding.
\newblock {\em ACM Trans. Graph.}, 41(4), jul 2022.
  \href{https://doi.org/10.1145/3528223.3530127}
{doi: {{%
10\hspace{.1pt}\discretionary{.}{%
}{.}\hspace{.4pt}1145\discretionary{/}{%
}{/}3528223\hspace{.1pt}\discretionary{.}{%
}{.}\hspace{.4pt}3530127}}}


\bibitem{NEURIPS19_pytorch}
A.~Paszke, S.~Gross, F.~Massa, A.~Lerer, J.~Bradbury, G.~Chanan, T.~Killeen,
  Z.~Lin, N.~Gimelshein, L.~Antiga, A.~Desmaison, A.~Kopf, E.~Yang, Z.~DeVito,
  M.~Raison, A.~Tejani, S.~Chilamkurthy, B.~Steiner, L.~Fang, J.~Bai, and
  S.~Chintala.
\newblock {PyTorch: An Imperative Style, High-Performance Deep Learning
  Library}.
\newblock In {\em Advances in Neural Information Processing Systems 32}, pp.
  8024--8035. 2019.

\bibitem{JHUTDB2}
E.~{Perlman}, R.~{Burns}, Y.~{Li}, and C.~{Meneveau}.
\newblock {Data Exploration of Turbulence Simulations using a Database
  Cluster}.
\newblock In {\em Proc. ACM/IEEE Conference on Supercomputing}, pp. 1--11,
  2007.

\bibitem{Reiser21_kilonerf}
C.~Reiser, S.~Peng, Y.~Liao, and A.~Geiger.
\newblock {KiloNeRF: Speeding up Neural Radiance Fields with Thousands of Tiny
  MLPs}.
\newblock {\em 2021 IEEE/CVF International Conference on Computer Vision
  (ICCV)}, pp. 14315--14325, 2021.

\bibitem{sitzmann19_siren}
V.~Sitzmann, J.~N. Martel, A.~W. Bergman, D.~B. Lindell, and G.~Wetzstein.
\newblock {Implicit Neural Representations with Periodic Activation Functions}.
\newblock In {\em Proc. NeurIPS}, 2020.

\bibitem{takikawa21_nglod}
T.~Takikawa, J.~Litalien, K.~Yin, K.~Kreis, C.~Loop, D.~Nowrouzezahrai,
  A.~Jacobson, M.~McGuire, and S.~Fidler.
\newblock Neural geometric level of detail: Real-time rendering with implicit
  {3D} shapes.
\newblock {\em arXiv preprint arXiv:2101.10994}, 2021.

\bibitem{tancik2022_blocknerf}
M.~Tancik, V.~Casser, X.~Yan, S.~Pradhan, B.~Mildenhall, P.~Srinivasan, J.~T.
  Barron, and H.~Kretzschmar.
\newblock {Block-NeRF}: Scalable large scene neural view synthesis.
\newblock {\em arXiv}, 2022.

\bibitem{tancik20_fourier}
M.~Tancik, P.~P. Srinivasan, B.~Mildenhall, S.~Fridovich-Keil, N.~Raghavan,
  U.~Singhal, R.~Ramamoorthi, J.~T. Barron, and R.~Ng.
\newblock Fourier features let networks learn high frequency functions in low
  dimensional domains.
\newblock {\em NeurIPS}, 2020.

\bibitem{treib12_cudacompress}
M.~Treib, K.~Burger, F.~Reichl, C.~Meneveau, A.~Szalay, and R.~Westermann.
\newblock {Turbulence Visualization at the Terascale on Desktop PCs}.
\newblock {\em {IEEE transactions on visualization and computer graphics}},
  18(12):2169--2177, 2012.

\bibitem{wang22_dl4scivis}
C.~Wang and J.~Han.
\newblock {DL4SciVis: A State-of-the-Art Survey on Deep Learning for Scientific
  Visualization}.
\newblock {\em IEEE transactions on visualization and computer graphics}, PP,
  April 2022. \href{https://doi.org/10.1109/tvcg.2022.3167896}
{doi: {{%
10\hspace{.1pt}\discretionary{.}{%
}{.}\hspace{.4pt}1109\discretionary{/}{%
}{/}tvcg\hspace{.1pt}\discretionary{.}{%
}{.}\hspace{.4pt}2022\hspace{.1pt}\discretionary{.}{%
}{.}\hspace{.4pt}3167896}}}


\bibitem{Weiss22_fvsrn}
S.~Weiss, P.~Hermüller, and R.~Westermann.
\newblock Fast neural representations for direct volume rendering.
\newblock {\em Computer Graphics Forum}, 41(6):196--211, 2022.
  \href{https://doi.org/10.1111/cgf.14578}
{doi: {{%
10\hspace{.1pt}\discretionary{.}{%
}{.}\hspace{.4pt}1111\discretionary{/}{%
}{/}cgf\hspace{.1pt}\discretionary{.}{%
}{.}\hspace{.4pt}14578}}}


\bibitem{wu2023_hyperinr}
Q.~Wu, D.~Bauer, Y.~Chen, and K.-L. Ma.
\newblock Hyperinr: A fast and predictive hypernetwork for implicit neural
  representations via knowledge distillation, 2023.

\bibitem{Wu22_ngpscivis}
Q.~Wu, D.~Bauer, M.~J. Doyle, and K.-L. Ma.
\newblock Instant neural representation for interactive volume rendering, 2022.
  \href{https://doi.org/10.48550/ARXIV.2207.11620}
{doi: {{%
10\hspace{.1pt}\discretionary{.}{%
}{.}\hspace{.4pt}48550\discretionary{/}{%
}{/}ARXIV\hspace{.1pt}\discretionary{.}{%
}{.}\hspace{.4pt}2207\hspace{.1pt}\discretionary{.}{%
}{.}\hspace{.4pt}11620}}}


\bibitem{yu21_plenoctrees}
A.~Yu, R.~Li, M.~Tancik, H.~Li, R.~Ng, and A.~Kanazawa.
\newblock {PlenOctrees} for real-time rendering of neural radiance fields.
\newblock In {\em ICCV}, 2021.

\bibitem{zhao21_sz3_3}
K.~Zhao, S.~Di, M.~Dmitriev, T.-L.~D. Tonellot, Z.~Chen, and F.~Cappello.
\newblock Optimizing error-bounded lossy compression for scientific data by
  dynamic spline interpolation.
\newblock In {\em 2021 IEEE 37th International Conference on Data Engineering
  (ICDE)}, pp. 1643--1654, 2021.
  \href{https://doi.org/10.1109/ICDE51399.2021.00145}
{doi: {{%
10\hspace{.1pt}\discretionary{.}{%
}{.}\hspace{.4pt}1109\discretionary{/}{%
}{/}ICDE51399\hspace{.1pt}\discretionary{.}{%
}{.}\hspace{.4pt}2021\hspace{.1pt}\discretionary{.}{%
}{.}\hspace{.4pt}00145}}}


\bibitem{SSIM}
{Zhou Wang}, A.~C. {Bovik}, H.~R. {Sheikh}, and E.~P. {Simoncelli}.
\newblock {Image quality assessment: from error visibility to structural
  similarity}.
\newblock {\em IEEE Transactions on Image Processing}, 13(4):600--612, 2004.

\end{thebibliography}


\appendix 

\end{document}